\documentclass{JINST}
\usepackage{graphicx}
\usepackage{epsfig}
\usepackage{amssymb,amsmath}
\usepackage{longtable}
\title{The linearity response of the Planck-LFI flight model receivers}

   \author{
        A.  Mennella$^a$\thanks{Corresponding author},
        F.  Villa$^b$,
        L.  Terenzi$^b$,
        F.  Cuttaia$^b$,
        P.  Battaglia$^c$,
        M.  Bersanelli$^a$,
        R.C. Butler$^b$,
        O.  D'Arcangelo$^d$,
        E. Artal$^e$,
        R. Davis$^f$,
        M.  Frailis$^g$,
        C.  Franceschet$^c$,
        S.  Galeotta$^g$,
        A.  Gregorio$^{h,g}$,
        N.  Hughes$^i$,
        P.  Jukkala$^i$,
        D. Kettle$^f$,
        V.-H. Kilpi\"{a}$^i$,
        M.  Laaninen$^j$,
        P.M.    Lapolla$^c$,
        R.  Leonardi$^k$,
        P.  Leutenegger$^c$,
        S.  Lowe$^f$,
        N.  Mandolesi$^b$,
        M.  Maris$^g$,
        P.  Meinhold$^k$,
        L.  Mendes$^l$,
        M.  Miccolis$^c$,
        G.  Morgante$^b$,
        N.  Roddis$^f$,
        M.  Sandri$^b$,
        R.  Silvestri$^c$,
        L.  Stringhetti$^b$,
        M.  Tomasi$^a$,
        J. Tuovinen$^m$,
        L.  Valenziano$^b$,
        A.  Zacchei$^g$,
        J. Varis$^m$,
        A.  Wilkinson$^f$ and
        A.  Zonca$^n$\\
    \llap{$^a$}Universit\`a degli Studi di Milano, Dipartimento di Fisica\\ 
    Via Celoria 16, 20133 Milano, Italy\\
    \llap{$^b$}INAF-IASF Bologna\\
    Via P.Gobetti 101, 40129 Bologna, Italy\\
    \llap{$^c$}Thales Alenia Space - Italia\\
    S.S. Padana Superiore 290, 20090 Vimodrone (Milano) - Italy\\
    \llap{$^d$}CNR, Istituto di Fisica del Plasma\\
    Via Roberto Cozzi 53, 20125 Milano, Italy\\
    \llap{$^e$}Departamento de Ingenier\`ia de Comunicaciones, Universidad de Cantabria\\
    Avenida de los Castros s/n. 39005 Santander, Spain\\
    \llap{$^f$}Jodrell Bank Centre for Astrophysics\\
    University of Manchester, M13 9PL, UK\\
    \llap{$^g$}INAF - Osservatorio Astronomico di Trieste\\
    Via Tiepolo 11, 34012 Trieste, Italy\\
    \llap{$^h$}Universit\`a degli Studi di Trieste, Dipartimenti di fisica\\
    Via Valerio 2, 34127 Trieste, Italy\\
    \llap{$^i$}DA-Design Oj\\
    Keskuskatu 29, FI-31600 Jokioinen, Finland\\
    \llap{$^j$}Ylinen Electronics Oy\\
    Teollisuustie 9A, FIN-02700 Kauniainen, Finland\\
    \llap{$^k$}University of California at Santa Barbara, Physics Department\\
    Santa Barbara CA 93106, USA\\
    \llap{$^l$} Planck Science Office, European Space Agency ESAC\\
    P.O. box 78 28691 Villanueva de la Ca\~{n}ada Madrid, Spain\\
    \llap{$^m$} MilliLab, VTT Technical Research Centre of Finland\\
    P.O. Box 1000, FI-02044 VTT, Finland\\
    \llap{$^n$}INAF-IASF Milano\\
    Via Bassini 15, 20133 Milano, Italy\\
    Email: \email{aniello.mennella@fisica.unimi.it}
}

   \abstract{
        In this paper we discuss the linearity response of the Planck-LFI receivers, with particular reference to signal compression measured on the 30 and 44 GHz channels. In the article we discuss the various sources of compression and present a model that accurately describes data measured during tests performed with individual radiomeric chains. After discussing test results we present the best parameter set representing the receiver response and discuss the impact of non linearity on in-flight calibration, which is shown to be negligible.
    }
    

\begin{document}

%

\def\aap{A\&A}%
\def\aapr{A\&A~Rev.}%
\def\aaps{A\&AS}%

\section{Introduction}
\label{sec:introduction}

  The Low Frequency Instrument (LFI) is an array of 22 coherent differential receivers at 30, 44 and 70 GHz on board the European Space Agency Planck satellite \cite{2009_LFI_cal_M1}. The LFI shares Planck telescope focal plane with the High Frequency Instrument (HFI), a bolometric array in the 100-857~GHz range cooled at 0.1~K. In 15 months of countinous measurements from the Lagrangian point L2, Planck will provide cosmic variance- and foreground-limited measurements of the Cosmic Microwave Background (CMB) by scanning the sky in almost great circles with a 1.5~m dual reflector aplanatic telescope \cite{2004_martin_planck_telescope,2002_villa_planck_telescope,2006_maris_planck_scanning_strategy,2005_dupac_planck_scanning_strategy}.

Best LFI noise performance is obtained with receivers based on High Electron Mobility Transistor (HEMT) amplifiers cryogenically cooled at 20~K by the Planck Sorption Cooler, a vibration-less hydrogen cooler providing more than 1~W of cooling power at 20~K. To optimise noise performance and cooling power the RF amplification is divided between a 20~K front-end unit and a $\sim$300~K back-end unit connected by composite waveguides \cite{2009_LFI_cal_M2}. 

The LFI has been calibrated and tested at different integration levels before testing individual receivers \cite{2009_LFI_cal_M4} and the whole receiver array \cite{2009_LFI_cal_M3}.

In this paper we discuss the Planck-LFI receivers response linearity. In particular we focus on the response of 30 and 44 GHz radiometers which show slight output compression extending over a wide range of input temperatures. This feature, that was discovered during the first tests on integrated receivers during the Qualification Model test campaign, affects the assessment of several performance parameters, like noise temperature, white noise sensitivity and noise effective bandwidth.

After a brief theoretical description of the basic receiver equations (see Section~\ref{sec:theory}) in Section~\ref{sec:linearity_receivers_response} we describe the characterisation of the receiver non linearity at 30 and 44~GHz, and discuss its cause. In the same section we also provide evidence for linear response of 70 GHz receivers in the temperature input range from $\sim$8~K to $\sim$40~K. We then discuss the impact on signal compression on ground calibration (Section~\ref{sec:impact_ground_calibration}) and on flight operations (Section~\ref{sec:impact_flight_operations}). This work is finally wrapped up and conclusions are provided in Section~\ref{sec:conclusions}. 

%
\section{Theory}
\label{sec:theory}


	

In each receiver assembly (also referred in this paper as Receiver Chain Assembly, RCA) the sky signal mirrored by the Planck telescope is received by a corrugated feed horn feeding an orthomode transducer (OMT) that splits the incoming wave into two perpendicularly polarised components. These propagate through two independent pseudo-correlation radiometers with HEMT (High Electron Mobility Transostor) amplifiers split between a cold ($\sim$20~K) and a warm ($\sim 300$~K) stage connected by composite waveguides \cite{2009_LFI_cal_O3}.

In this section we briefly introduce the LFI pseudo-correlation receiver theory and design and then we discuss in more detail the output response in case of linear and compressed behaviour. Further details about the LFI design can be found in \cite{2009_LFI_cal_M1, 2009_LFI_cal_M2,seiffert02, mennella03}.

\subsection{Receiver design}
\label{sec:receiver_design}

%
%

\begin{figure}[h!]
      \begin{center}
	 \includegraphics[width=12cm]{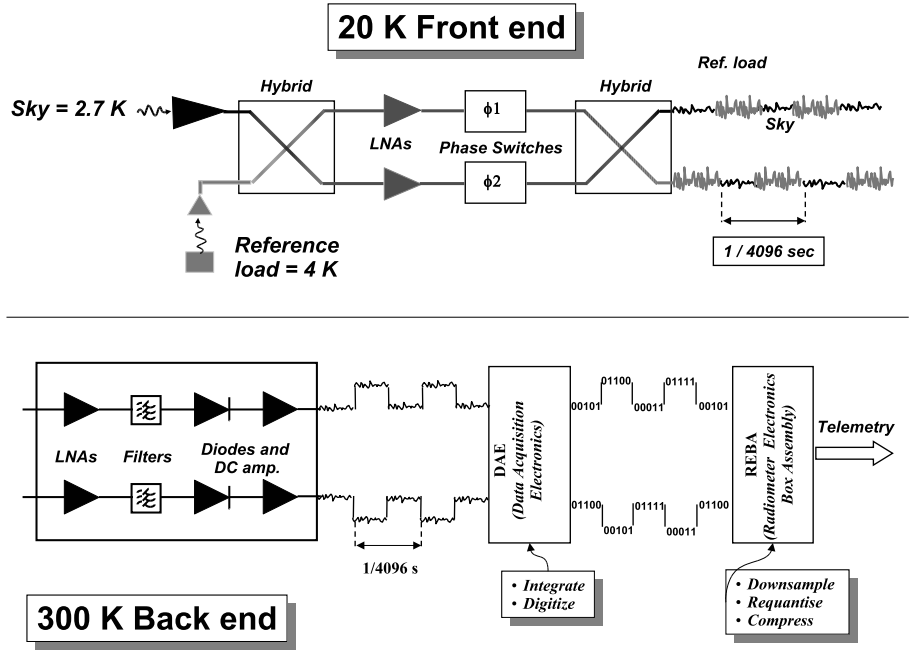}
      \end{center}
      \caption{Schematic of the LFI receivers pseudo correlation architecture}
      \label{fig:lfi_pseudo_correlation_schematic}
   \end{figure}

   A schematic of the LFI pseudo correlation receiver is shown in Figure~\ref{fig:lfi_pseudo_correlation_schematic}. In each radiometer the sky signal and a stable reference load at $\sim$4~K \cite{2009_LFI_cal_R1} are coupled to cryogenic low-noise HEMT amplifiers via a 180$^\circ$ hybrid. A phase shift oscillating between 0 and 180$^\circ$ at a frequency of 4096~Hz is then applied to one of the two signals. A second phase switch is present for symmetry on the second radiometer leg but it does not introduce any phase shift. A second 180$^\circ$ hybrid coupler recombines the signals so that the output is a sequence of sky-load outputs alternating at twice the frequency of the phase switch.

   In the back-end of each radiometer (see bottom part of Figure~\ref{fig:lfi_pseudo_correlation_schematic}) the RF signals are further amplified, filtered by a low-pass filter and then detected. After detection the sky and reference load signals are integrated and digitised in 14-bit integers by the LFI Digital Acquisition Electronics (DAE) box. Further binning and software quantisation is performed in the Radiometer Electronics Box Assembly (REBA), a digital processing unit that manages telemetry packet production from the raw instrument digital output. Further details about REBA and digital signal processing are described in \cite{2009_LFI_cal_E1} and \cite{2009_LFI_cal_D2}.

    The various RCAs are tagged with labels from LFI18 to LFI28 (see Table~\ref{tab:rca_id_correspondence}); each of the two radiometers connected to the two OMT arms are be labelled as M-0 (\textit{main} OMT arm) and S-1 (\textit{side} OMT arm, see \cite{2009_LFI_cal_O2}) while the two output detectors from each radiometer are be labelled as 0 and 1. Therefore with the label LFI18S-1, for example, we indicate the radiometer S of the RCA LFI18, and with the label LFI24M-01 we indicate detector 1 of radiometer M-0 in RCA LFI24. 
    
    \begin{table}[h!]
        \caption{Correspondence between receiver centre frequency and RCA label}
        \label{tab:rca_id_correspondence}
        \begin{center}
            \begin{tabular}{l l}
                \hline
                \mbox{} & \\
                70~GHz & LFI18 through LFI23\\
                44~GHz & LFI24, LFI25 and LFI26\\
                30~GHz & LFI27 and LFI28\\
                \mbox{} & \\
                \hline
            \end{tabular}
        \end{center}
    \end{table}

\subsection{Signal output}
\label{sec:signal_output}
 
   If the receiver isolation is perfect (i.e. if the sky and reference load signals are completely separated after the second hybrid) the relationship linking $T_{\rm in}$ to $V_{\rm out}$ can be written as:

   \begin{equation}
      V_{\rm out} = G(T_{\rm in}, T_{\rm noise})\times\left(T_{\rm in}+T_{\rm noise}\right),
      \label{eq:vout}
   \end{equation}
   where $T_{\rm in}$ refers to either $T_{\rm sky}$ or $T_{\rm ref}$, $V_{\rm out}$ is the corresponding voltage output, $T_{\rm noise}$ is the noise temperature and $G(T_{\rm in}, T_{\rm noise})$ is the calibration factor that, in general, may depend on the input and noise temperatures.


   In case of a linear response the calibration factor is a constant so that $G(T_{\rm in}, T_{\rm noise})\equiv G_0$. In Planck-LFI all the 70 GHz have proved very linear over a wide span of temperature inputs, ranging from $\sim 8$~K to $\sim 40$~K, while receivers at 30 and 44 GHz, instead, have shown slight compression that called for the development of a non linear response model. 

    In the following of this section we provide an overview of the response model from the analytical point of view, while in Section~\ref{sec:linearity_receivers_response} we discuss the source of the non linearity, showing that it is linked to compression in the back-end RF amplifiers and in the detector diode. 

    The parametrisation has been chosen following the work described in \cite{1989_daywitt_nonlinear_equations}. According to this work compression in the back-end of a radiometric receiver is modelled with a variable gain (i.e. that depends on the input power) with the analytical form described in Eq.~(\ref{eq:non_linearity_parametrisation}):

   \begin{eqnarray}
   \label{eq:non_linearity_parametrisation}
   \mbox{FEM} &=& \left\{ \begin{array}{ll}
         \mbox{Gain} = G^{\rm FEM}\\
         \mbox{Noise} = T_{\rm noise}^{\rm FEM}\end{array}
         \right. \nonumber\\
   &&\mbox{}\\
    \mbox{BEM} &=& \left\{ \begin{array}{ll}
         \mbox{Gain} = G^{\rm BEM} = \frac{G_0^{\rm BEM}}{1+b\cdot G_0^{\rm BEM}\cdot p}\\
         \mbox{Noise} = T_{\rm noise}^{\rm BEM},\end{array}
         \right.\nonumber
   \end{eqnarray}
    where FEM stands for \textit{front-end module}, $p$ is the power entering the BEM and $b$ is a parameter defining the BEM non linearity. This relationship is simple, correctly describes the limits of linear response ($b=0$) and infinite compression ($b=\infty$) and fits very well the radiometric response curves (see plots in Appendix~\ref{sec:best_fits}). This parametrisation therefore constituted our base model to characterise the radiometric voltage output response.
        

   The power entering the BEM (we neglect waveguide attenuation which may be included in the FEM parameters) is:
   \begin{equation}
       p = k \beta G_0^{\rm FEM} \left(T_{\rm in} + \tilde T_{\rm noise}\right),
      \label{eq:bem_input_power}
   \end{equation}
   where $\beta$ is the bandwidth, $k$ the Boltzmann constant, and $\tilde T_{\rm noise} = T_{\rm noise}^{\rm FEM} + \frac{T_{\rm noise}^{\rm WG}}{G_0^{\rm FEM}}$.
   So at the output of the BEM we have (the diode constant is considered inside the BEM gain):

   \begin{eqnarray}
      V_{\rm out} &=& k\beta G_0^{\rm FEM}\frac{G_0^{\rm BEM}\left(T_{\rm in}+T_{\rm noise}\right)}{1+b k \beta G_0^{\rm FEM} G_0^{\rm BEM}\left(T_{\rm in}+T_{\rm noise}\right)}
      = \frac{G_0\left(T_{\rm in}+T_{\rm noise}\right)}{1+b G_0\left(T_{\rm in}+T_{\rm noise}\right)}\nonumber\\
      \mbox{}\\
      G_0 &=& G_0^{\rm FEM} G_0^{\rm BEM} k \beta\nonumber
      \label{eq:vout_full}
   \end{eqnarray}
   which can be written in the following compact form:

   \begin{eqnarray}
      &&V_{\rm out} = G_{\rm tot}\left(T_{\rm in}+T_{\rm noise}\right)\nonumber\\
      &&G_{\rm tot} =  \frac{G_0}{1+b G_0\left(T_{\rm in}+T_{\rm noise}\right)}
      \label{eq:vout_compact}
   \end{eqnarray}

   We see from Eq.~(\ref{eq:vout_compact}) that the in the case of $b=0$ it reduces to the classical linear equation, whereas if $b\neq 0$ the equation tells us that the receiver gain is not constant but dependent on the input and noise temperatures coupled with the non-linearity parameter.


\section{Linearity in LFI receivers response}
\label{sec:linearity_receivers_response}

  In this section we discuss the various potential sources of compression in the LFI receivers. In particular we show how compressed behaviour was found in the 30 and 44 GHz receivers and that it was determined essentially by the back-end RF amplifiers and diodes. We also show that the 70~GHz receivers always provided a linear response in the tested input signal range.

\subsection{Sources of compression in LFI receivers}
\label{sec:sources_of_compression}
 
The linearity in a microwave receiver depends on the response of its individual components: radio-frequency amplifiers, detector diode and back-end analog electronics. 


The main potential sources of compression in the LFI receivers are represented by the RF amplifiers in the front-end and back-end modules and the back-end square-law detector. Let us now estimate the input power at the various stages (FEM amplification, BEM amplification and detector) expected during nominal operations, i.e. observing an input temperature of $\sim$2.7~K. The input power at a given stage in the radiometric chain can be calculated from the following relationship:

\begin{equation}
    P_{\rm in} = k\beta G\left(T_{\rm in}+T_{\rm noise}\right)
    \label{eq:input_power}
\end{equation}
where $T_{\rm in}$ is the input antenna temperature, $G$ and $T_{\rm noise}$ are gain and noise temperature of the radiometric chain before the stage considered in the calculation, $\beta$ is the bandwidth and $k$ the Boltzmann constant. Table~\ref{tab:input_powers} summarises estimates of the input power at the various receiver stages based on typical gain, noise temperature and bandwidth values.

\begin{table}[h!]
    \caption{Typical input power in dBm to the various receiver stages. The calculation has been performed  
    using the following typical parameters: $G^{\rm FEM} = 30$~dB, $G^{\rm BEM} = 35$~dB, $\beta = 20\%$ of the centre frequency, $T_{\rm noise} = 10$~K at 30~GHz, 16~K at 44~GHz and 30~K at 70~GHz.}
    
    \label{tab:input_powers}
    \begin{center}
        \begin{tabular}{l c c c}
            \hline
            \hline
                &30 GHz &44 GHz &70 GHz\\
            \hline
            Front-end   &-98    &-97    &-96\\
            Back-end    &-60    &-57    &-52\\
            Diode       &-25    &-22    &-17\\
            \hline
        \end{tabular}
    \end{center}
\end{table}

From Table~\ref{tab:input_powers} it is apparent that the input power to front-end amplifiers is extremely low, and very far from the typical compression levels of HEMT devices. Back-end RF amplifiers and, especially, detector diodes, receive a much higher input power so that they can be a source of non linear response. 

In particular this showed to be the case for 30 and 44 GHz back-end modules as discussed in detail in Section~\ref{sec:back_end_characterisation}. It must be noticed that input power received by back-end RF amplifier and detector diodes is actually higher in 70~GHz receivers compared to 30 and 44~GHz, which appears to be in  contradiction with the observed behaviour. We must underline, however, that 30 and 44~GHz BEMs components are different compared to 70~GHz BEMs; in particular RF amplifiers in low frequency BEMs are based on GaAs MMIC devices while in 70~GHz BEMs InP MMIC devices have been used. Further details about BEMs components and response can be found in \cite{2009_LFI_cal_R9,2009_LFI_cal_R10}.

\subsection{Characterisation of non linearity}
\label{sec:non_linearity_characterisation}

  \subsubsection{Characterisation of receiver response}
  \label{sec:characterisation_of_receiver_response}
  
    
    The linearity response of the LFI receivers has been derived by measuring, for each output channel, the radiometer voltage output, $V_{\rm out}$, at various input temperatures of the reference loads, $T_{\rm in}$, ranging from $\sim$8~K to $\sim$30~K. Then the linearity parameter $b$ can be determined by fitting the acquired data $V_{\rm out}^j(T_{\rm in}^j)$ with Eq.~(\ref{eq:vout_compact}), where the fitting parameters are $G_0$, $T_{\rm noise}$ and $b$ (see Section~\ref{sec:noise_t_calibration_const}).

    We have also charecterised linearity with a different and somewhat simpler approach, that avoids a three-parameters fit and allows to define a normalised non-linearity parameter that is independent of the receiver characteristics provided that the temperature range over which linearity is characterised is approximately the same for all detectors. This parameter has been calculated as follows:

    \begin{itemize}
        \item remove the average from the measured input temperature and output voltage, i.e. calculate $\tilde V_{\rm out}^j = V_{\rm out}^j - \langle V_{\rm out}\rangle_j$ and $\tilde T_{\rm in}^j = T_{\rm in}^j - \langle T_{\rm in}\rangle_j$;
        \item fit the $\tilde V_{\rm out}^j(\tilde T_{\rm in}^j)$ data with a straight line calculating a slope $s$;
        \item multiply the voltage outputs by the calculated slope, i.e. calculate $\bar T_{\rm out}^j = s\times\tilde V_{\rm out}^j$;
        \item calculate $L = \sum \left( \bar T_{\rm out}^j - \tilde T_{\rm in}^j \right)^2$.
    \end{itemize}

    In case of a perfect linear response then $\bar T_{\rm out}^j = \tilde T_{\rm in}^j$ (i.e. measured points, after normalisation, lie on a $y=x$ line) and $L=0$. The parameter $L$, therefore, provides a measure of deviation from linearity. 

\begin{figure}[h!]
        \begin{center}
            \begin{tabular}{c|c}
            \includegraphics[width = 6.6cm]{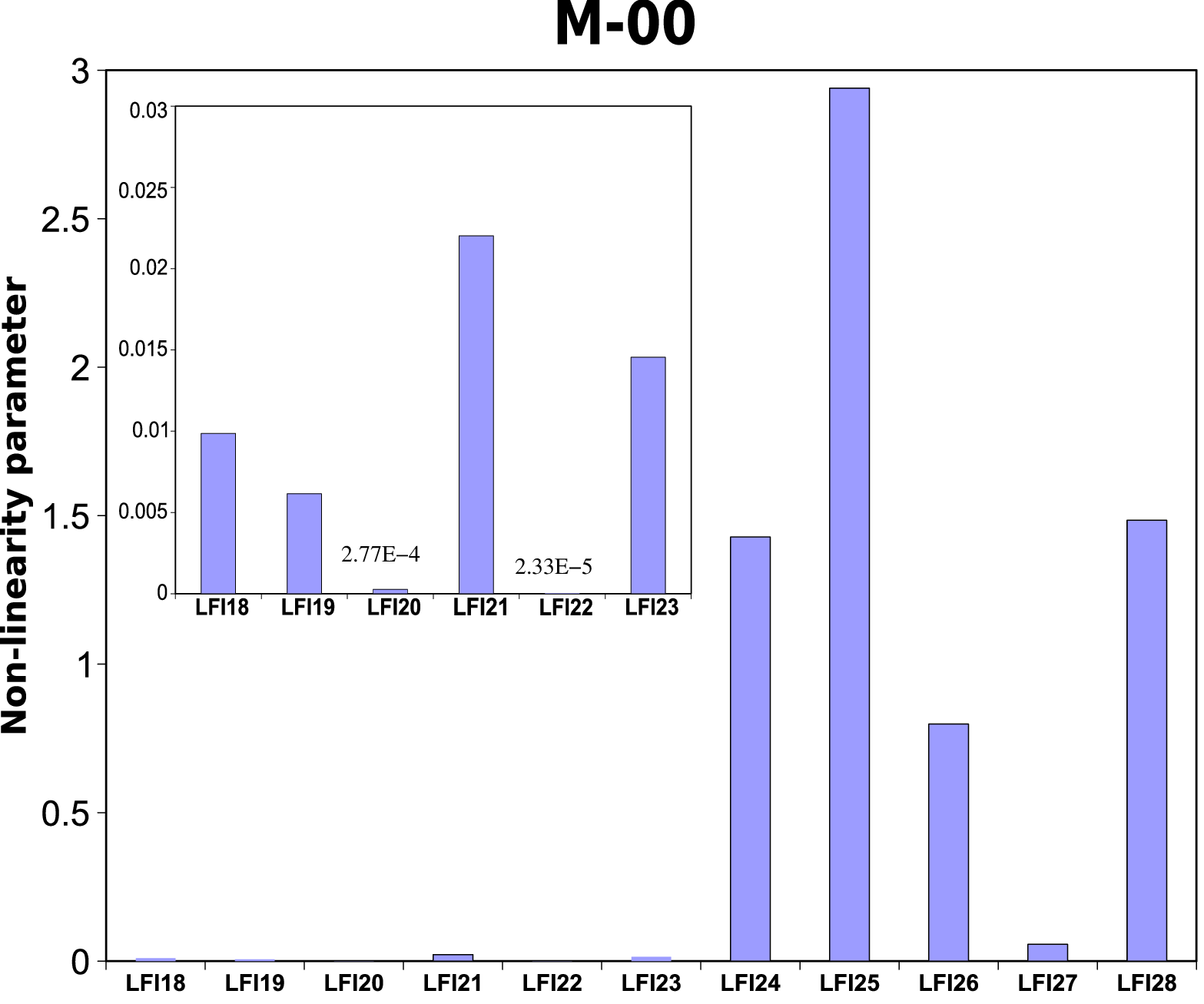} &
            \includegraphics[width = 6.8cm]{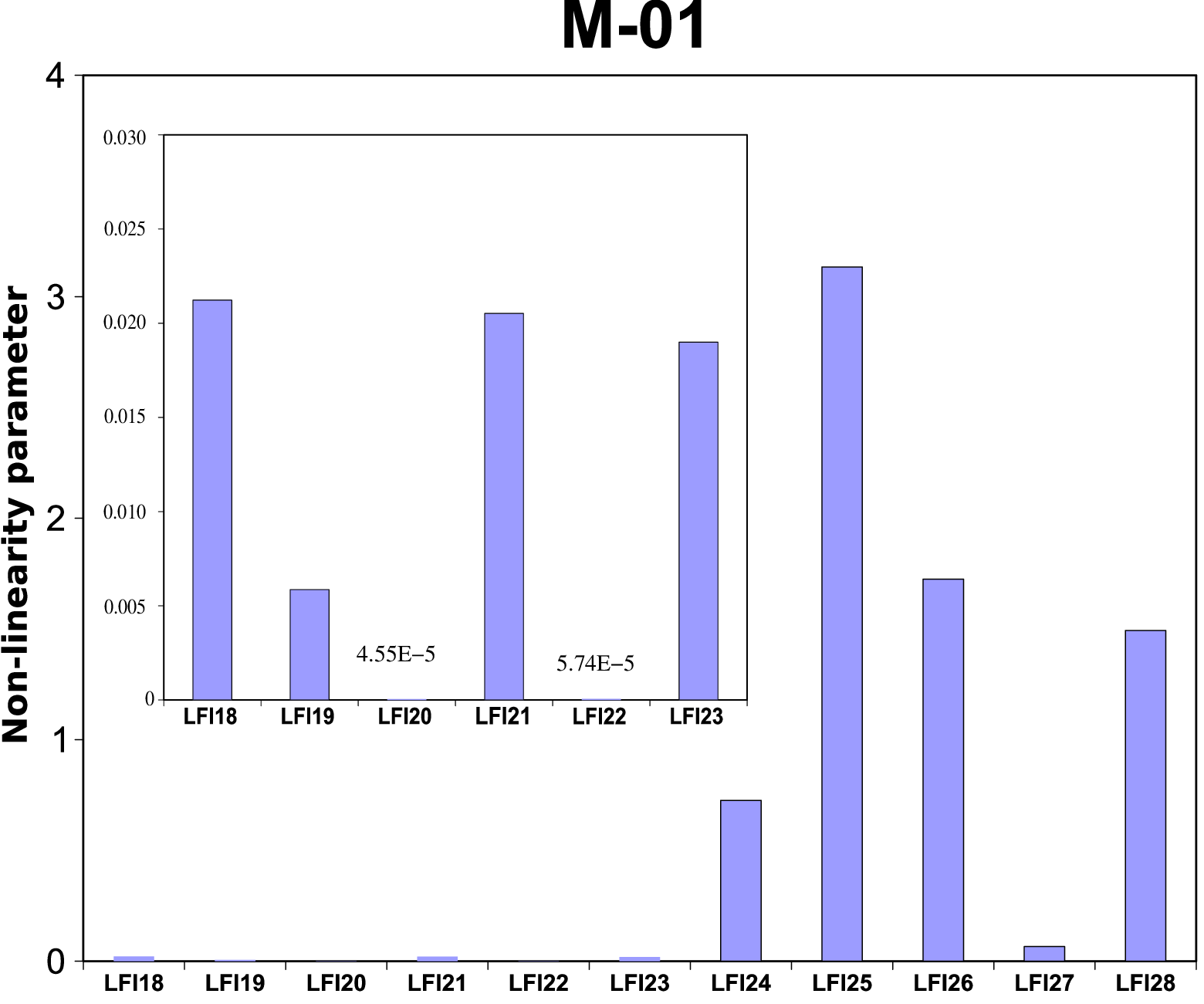} \\
                \hline
            \includegraphics[width=6.8cm]{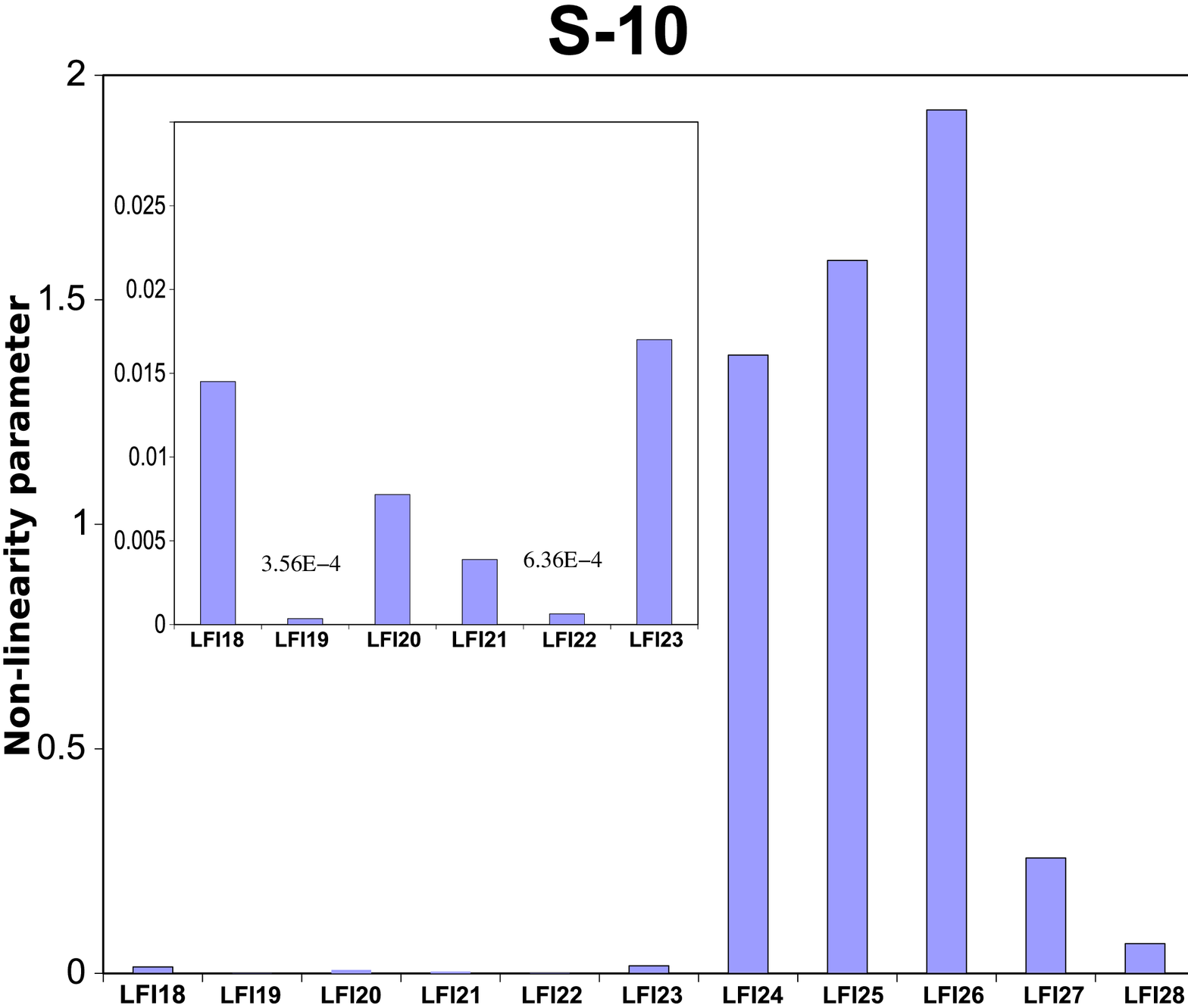} &
            \includegraphics[width=7cm]{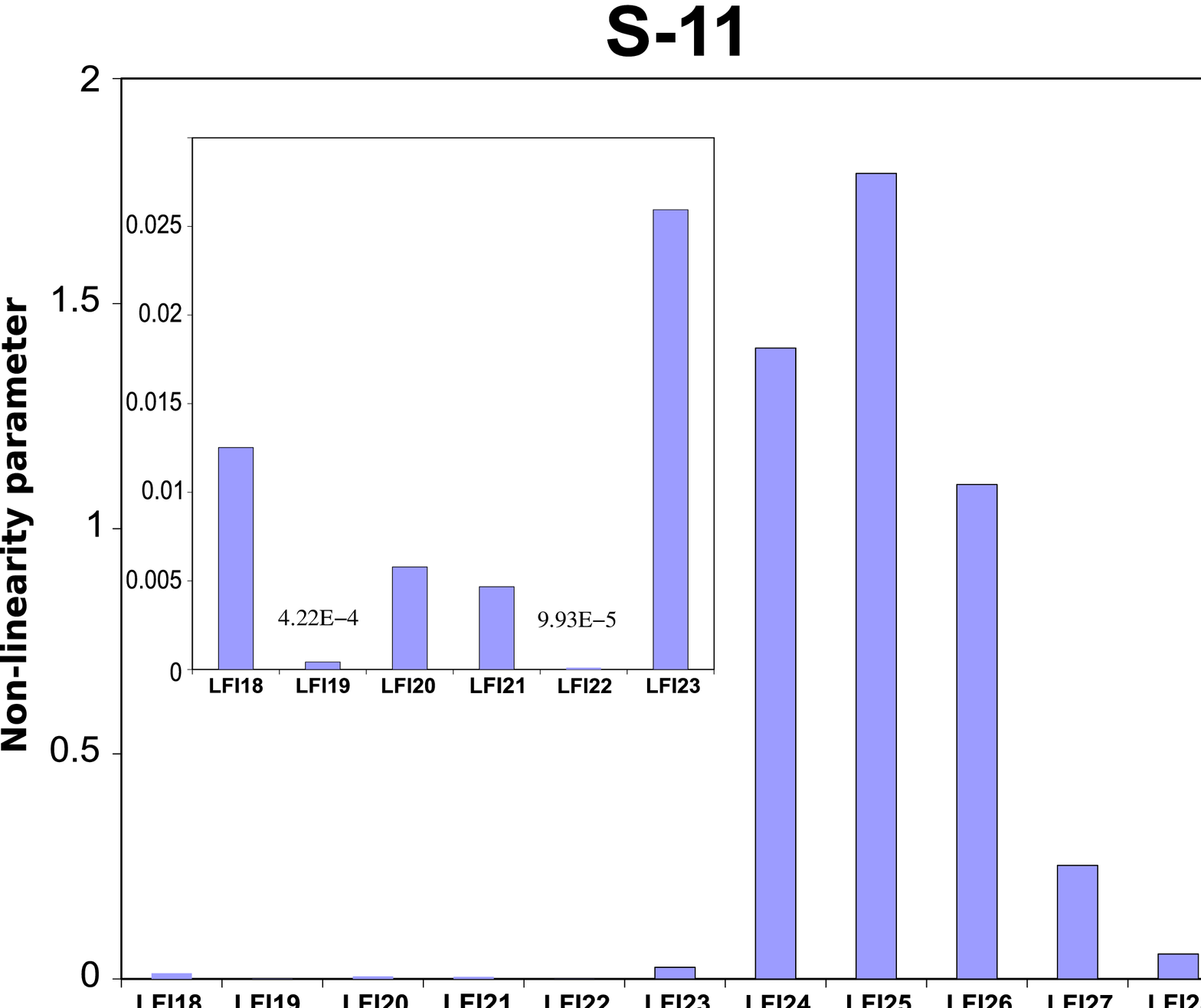} 
            \end{tabular}
        \end{center}
        \caption{Non linearity parameters for all LFI channels. In each plot a small inset provides a zoom on the 70 GHz non linearity parameters on an expanded scale.}
        \label{fig:non_linearity_parameters_all}
    \end{figure}

    A comprehensive view of the values of $L$ for all detectors (calculated in a input temperature range of the reference load between $\sim$9~K and $\sim$30~K) is provided in Figure~\ref{fig:non_linearity_parameters_all}. From the figure it is apparent that 70~GHz detectors are extremely linear while 30 and, especially, 44~GHz detectors show significant non-linearity.

\begin{figure}[h!]
        \begin{center}
            \includegraphics[width=10cm]{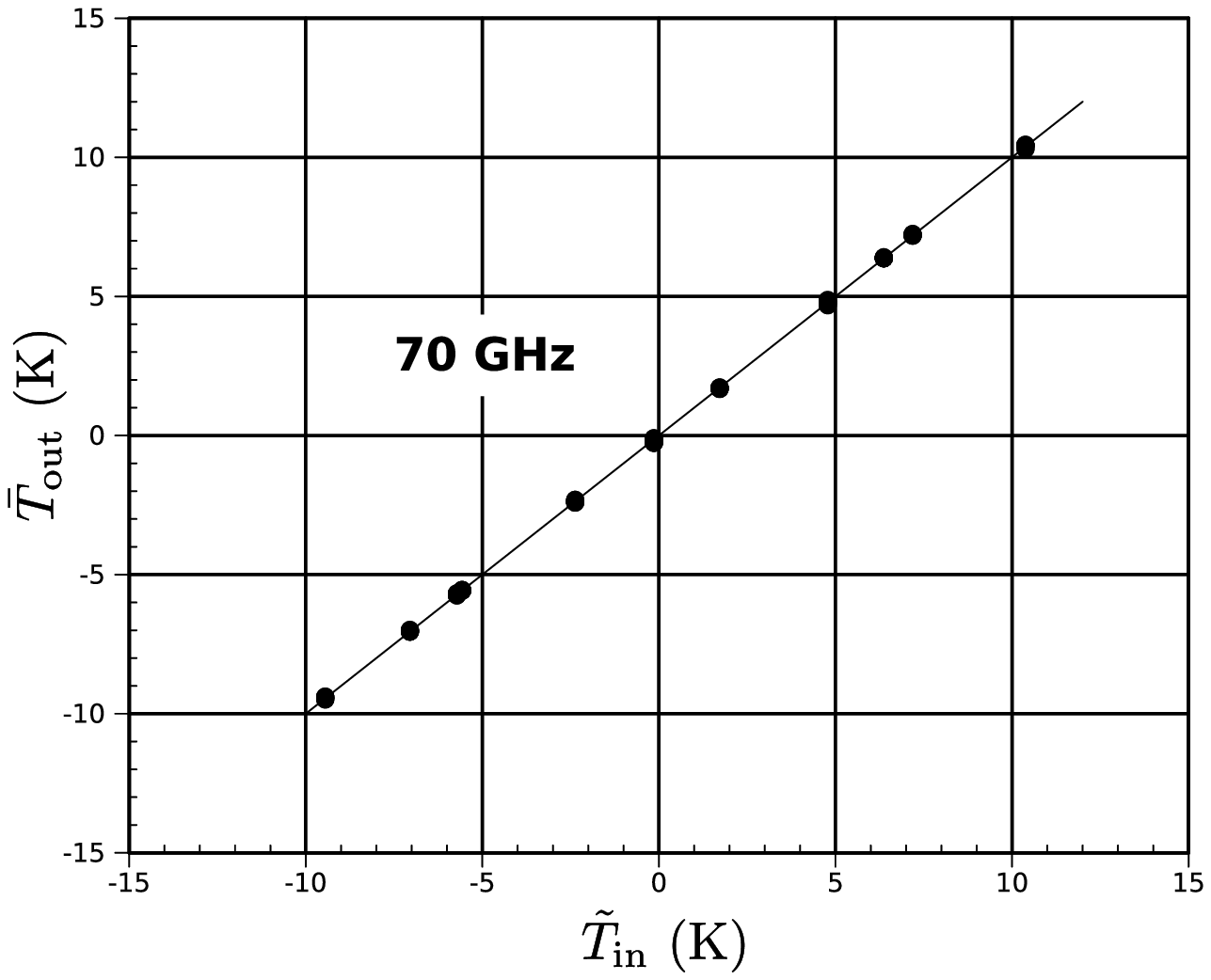}
            \caption{Normalised output response from all 70 GHz detectors.}
        \end{center}
        \label{fig:70_ghz_linearity}
    \end{figure}

    In Figure~\ref{fig:70_ghz_linearity} we show a comprehensive plot of the normalised receiver response from all 24 70 GHz detectors. Notice that the measured points almost perfectly lie on the $y = x$ line. Furthermore it may be noticed that the plot appears to display much less points than exptected from 24 detectors; this is because for each normalised temperature, $\tilde T_{\rm in}^j$, the normalised voltage values from the various detectors essentially overlap.

\begin{figure}[h!]
        \centering
        \includegraphics[width=7.5cm]{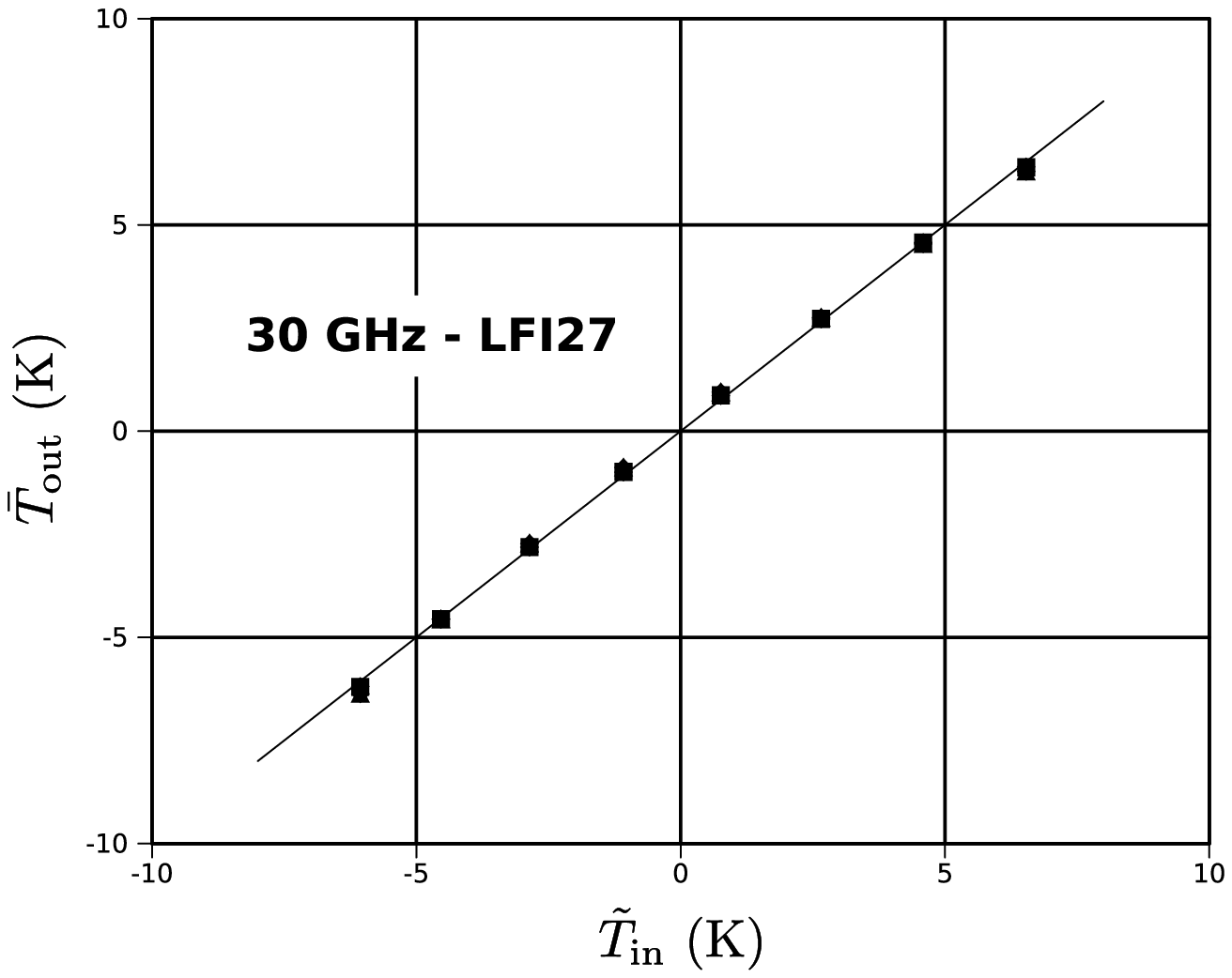} 
        \includegraphics[width=7.5cm]{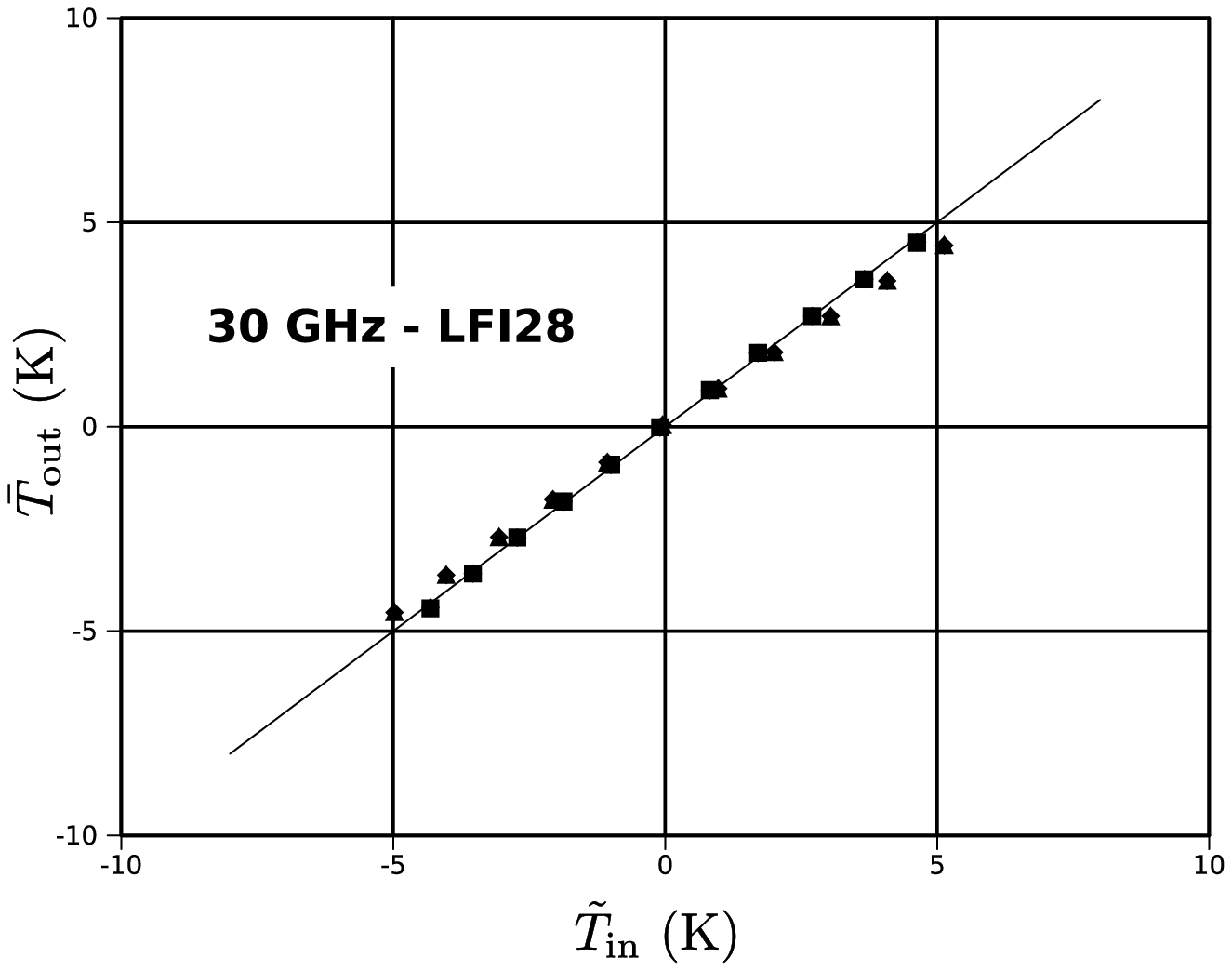} 
        \caption{Normalised output response from 30 GHz detectors. Each plot represents data from the 4 detectors of each RCA.}
        \label{fig:30_ghz_linearity}
    \end{figure}

\begin{figure}[h!]
        \centering
        \includegraphics[width=7.5cm]{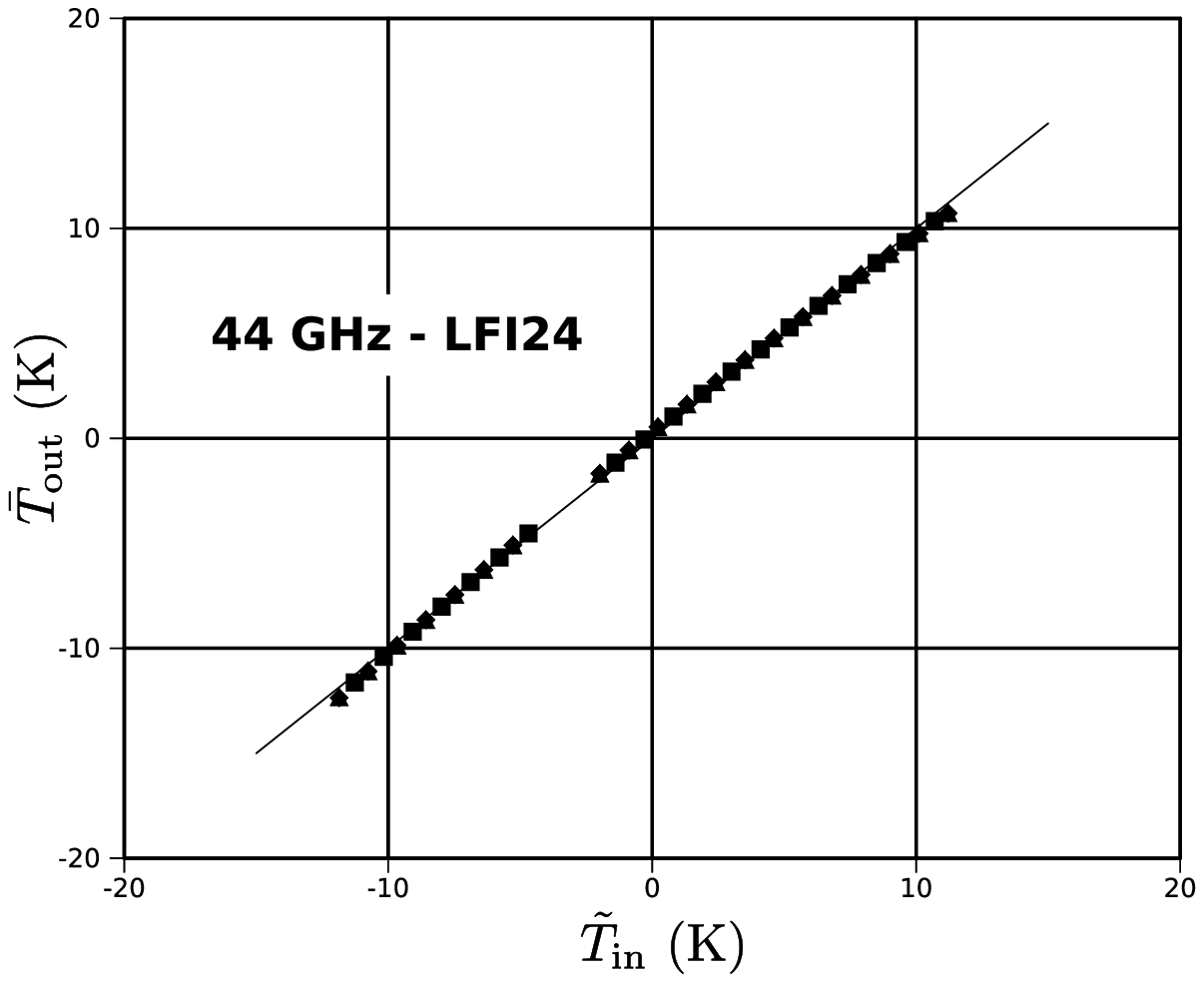}
        \includegraphics[width=7.5cm]{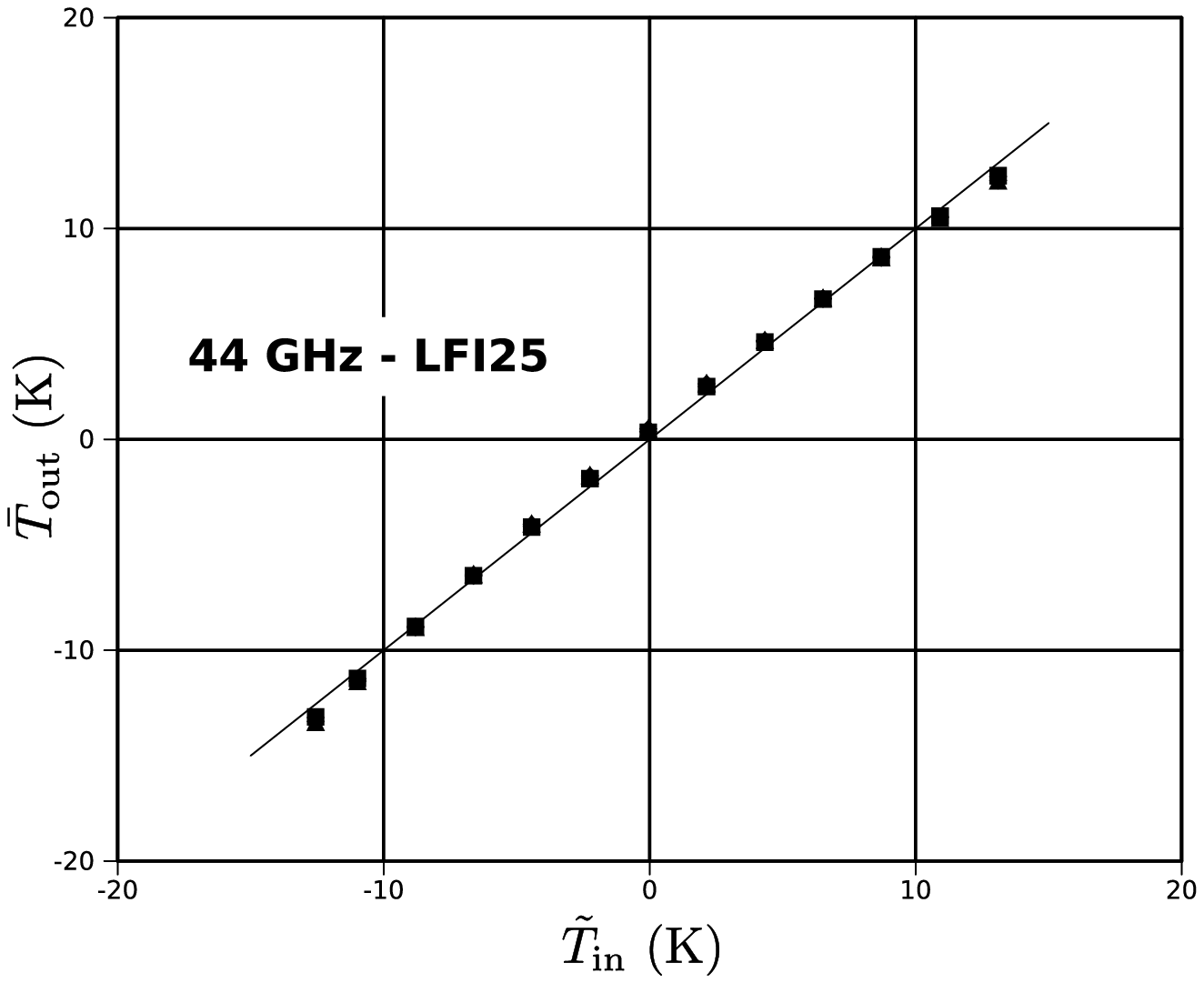} \\
        \includegraphics[width=7.5cm]{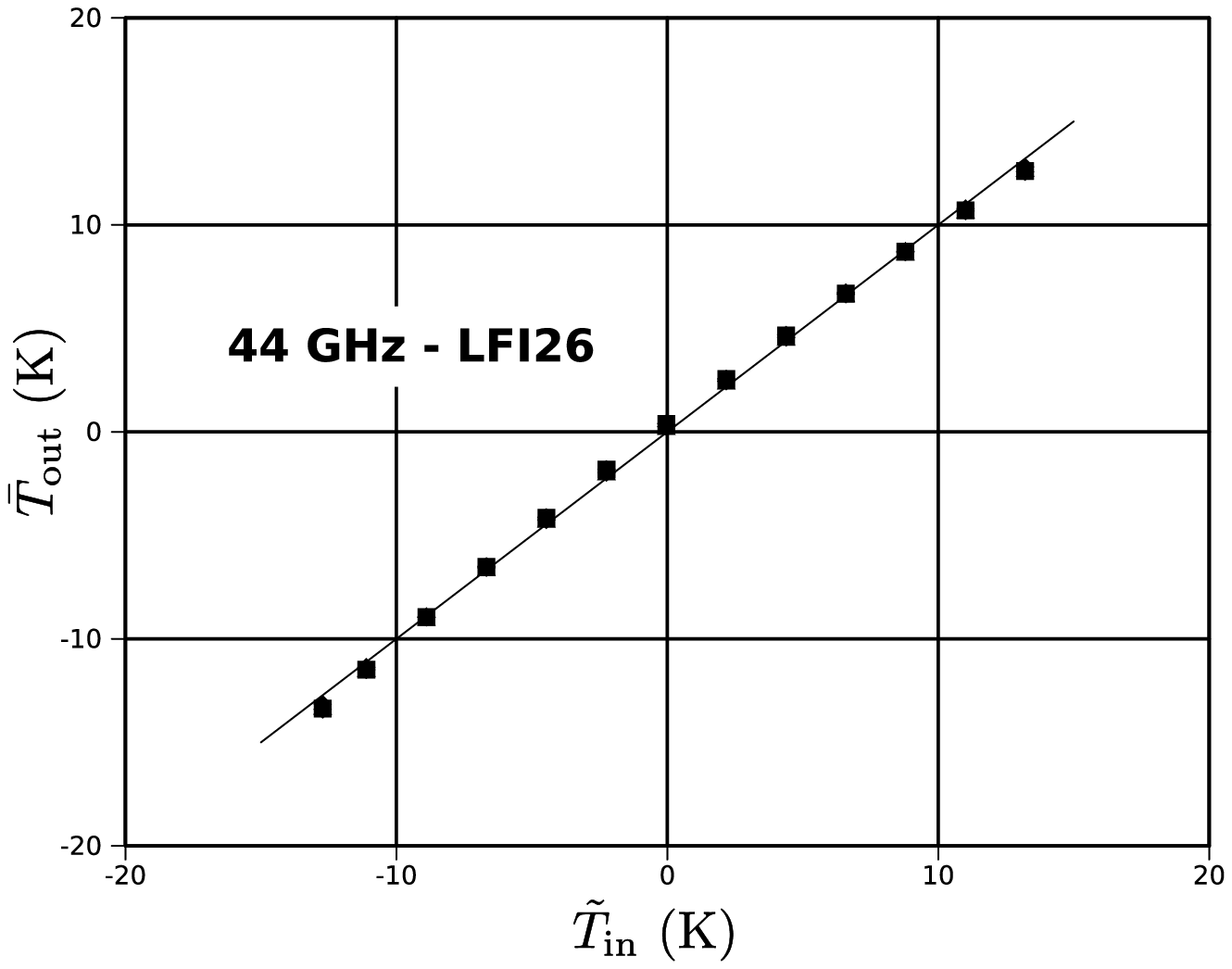} 
        \caption{Normalised output response from 44 GHz detectors. Each plot represents data from the 4 detectors of each RCA.}
        \label{fig:44_ghz_linearity}
    \end{figure}

    In Figures~\ref{fig:30_ghz_linearity} and \ref{fig:44_ghz_linearity} the same plot clearly shows significant deviations from linearity, especially for the 44~GHz receivers. Because, in this case, non linearity varies among the various detectors and overplotting all the data in each frequency channel would make the plots difficult to read, we have plotted the normalised voltage output for each RCA in different graphs.

    Deviation from linearity in the 30 and 44 GHz receivers, instead, is caused by signal compression caused by back-end RF amplifiers and diodes in presence of a broad-band signal. This is discussed in more detail in the next section, where we present some tests that were performed on two back-end units at 44~GHz and that provided the best characterisation of the signal compression in a very wide input power range.

  \subsubsection{Characterisation of back-end response}
  \label{sec:back_end_characterisation}
  
\begin{figure}[h!]
        \begin{center}
            \includegraphics[width=10cm]{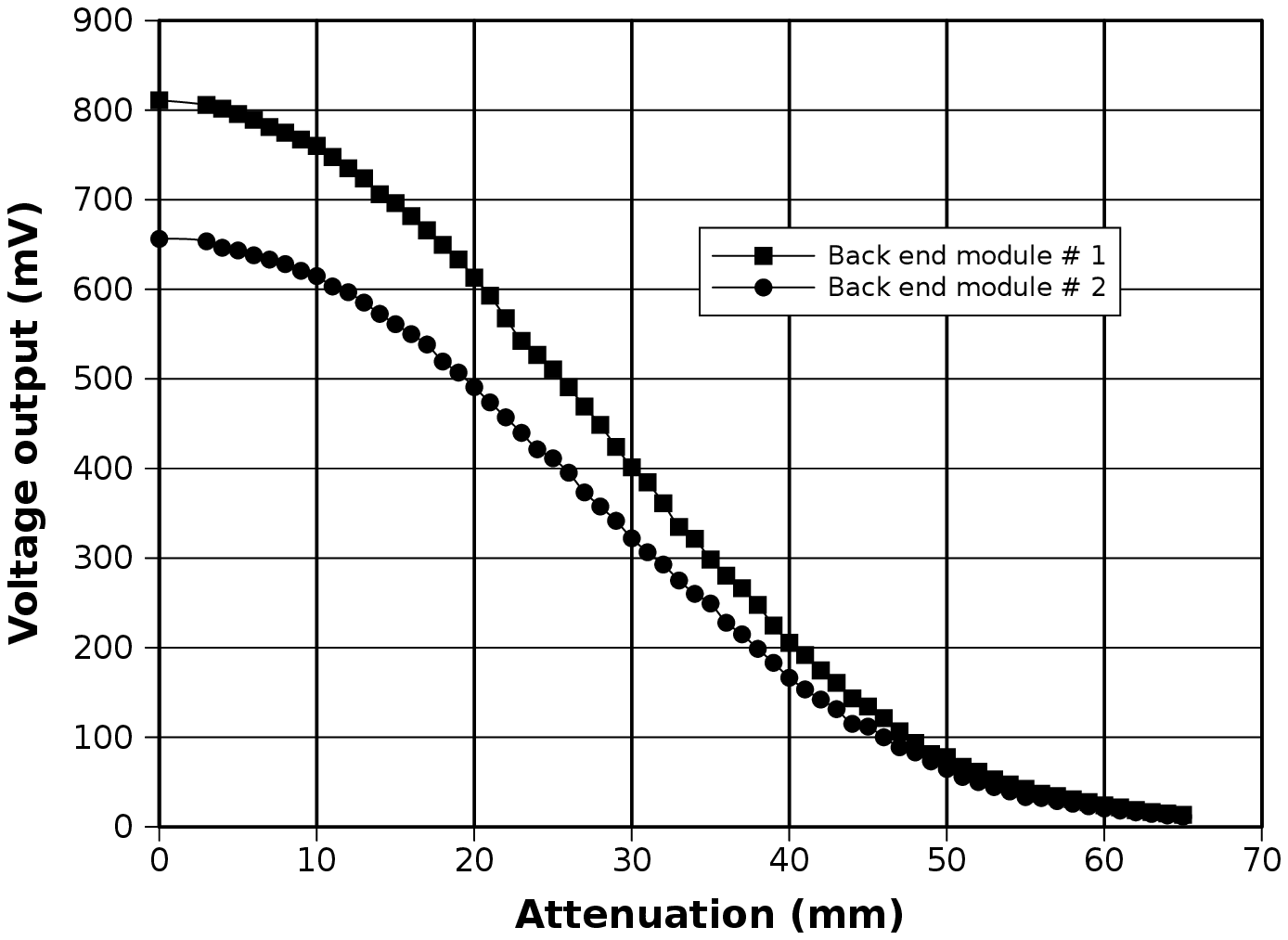}
        \end{center}
        \caption{Output voltage (with offset removed) from both back end modules as a function of the raw attenuation in mm.}
        \label{fig:raw_attenuation_curves}
    \end{figure}

    A set of tests have been performed on two back end modules at 44 GHz with the aim to identify the source of compression (RF amplifier or diode). The test was performed by observing with the receiver a sky and reference signal at $\sim$25~K and $\sim$18~K, respectively, and varying the input power to the back end with a variable power attenuator placed between the front and back end and coupled to a multimeter. In Figure~\ref{fig:raw_attenuation_curves} we show the output (after offset removal) from the two back ends as a function of the attenuator position in millimetres.
    
    The next step has been to calculate the input power to the back end module as a function of the attenuator position. This has been done using two independent methods, i.e.: (i) using a power meter to record the integrated signal reaching the back-end and (ii) using a noise figure meter to measure the input signal level versus frequency.
    
\begin{figure}[h!]
        \begin{center}
            \includegraphics[width=10cm]{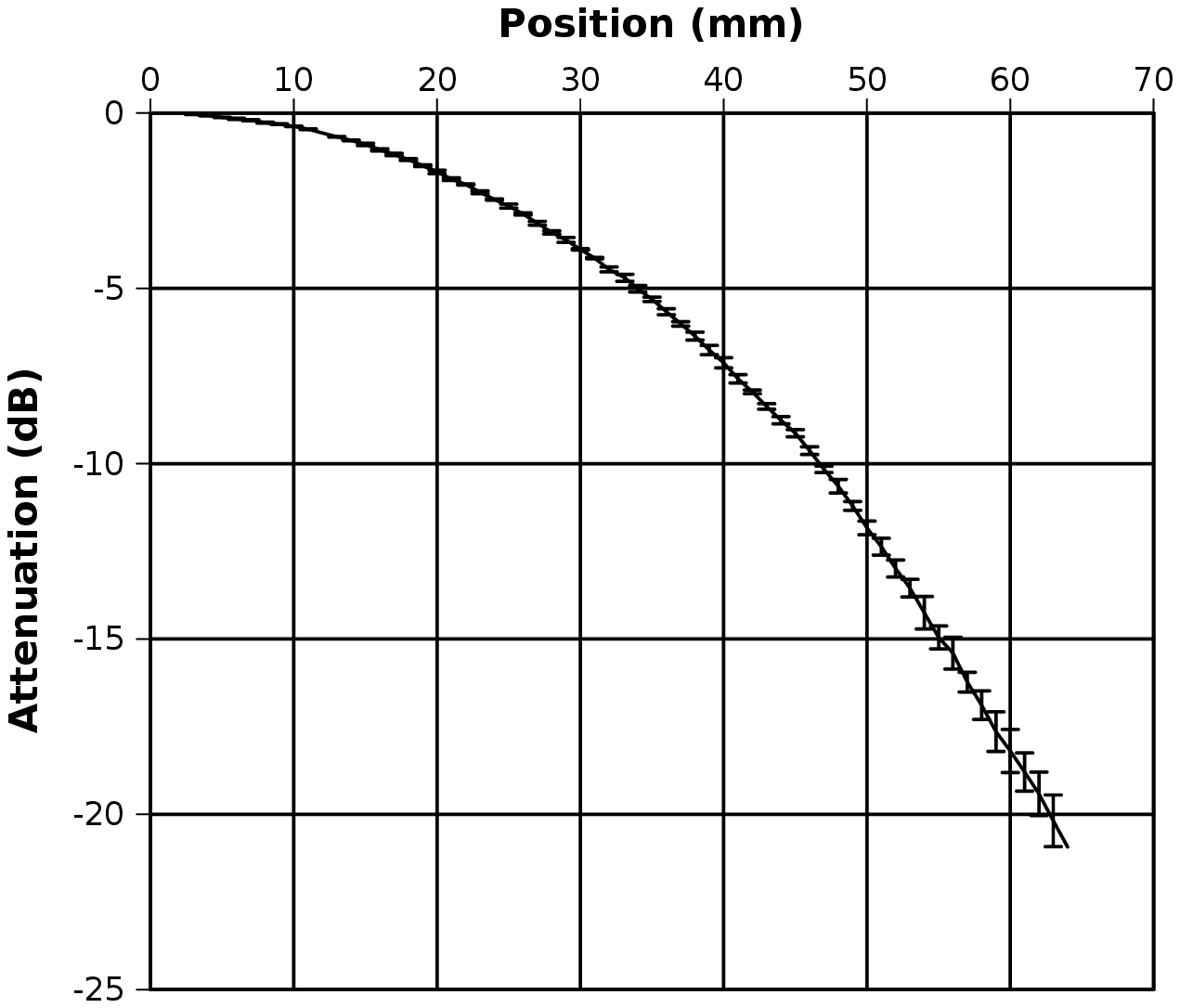}
        \end{center}    
        \caption{Attenuation as a function of attenuator position: curve and error bars are the result of three independent measurements.}
        \label{fig:attenuation_power_meter}
    \end{figure}

    \paragraph{Attenuation curves using a power meter.} 
    A power meter with a dynamic range up to -70 dBm was previously calibrated using its internal reference source and used to measure signals from the front-end module attenuated down to -21 dB.
    Three independent measurements taken in different days and configurations showed good repeatability, as shown in Figure~\ref{fig:attenuation_power_meter}.

\begin{figure}[h!]
        \begin{center}
            \includegraphics[width=10cm]{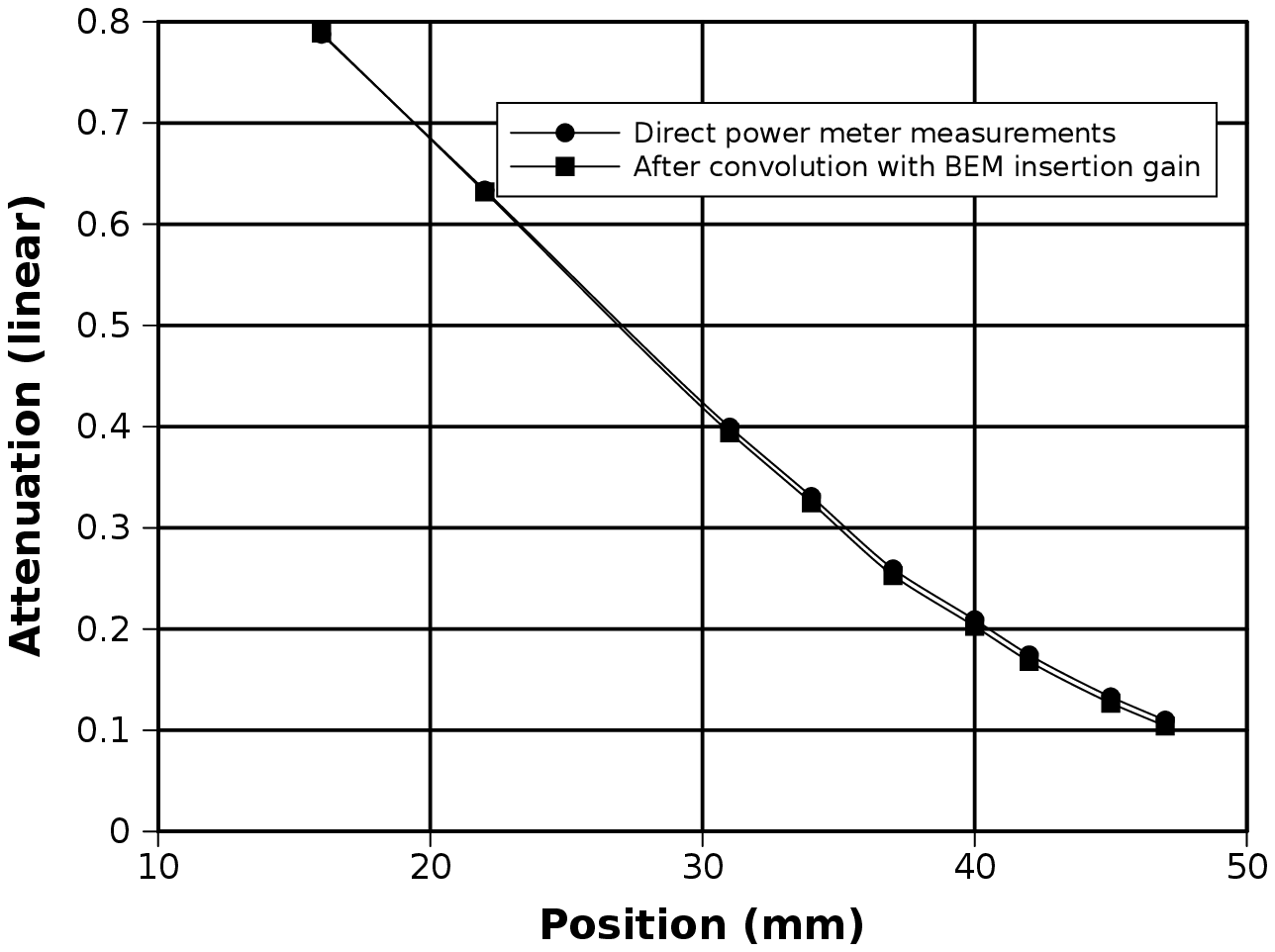}
        \end{center}
        \caption{Comparison of attenuation curves from the direct power meter measurements and from the same measurements after convolution with the RF insertion gain of a similar back-end module. Differences are very small.}
        \label{fig:attenuation_power_meter_comparison_insertion_gain}
    \end{figure}

    It is worth noting that the curve in Figure~\ref{fig:attenuation_power_meter} is an approximation of the effective attenuation, because it should be calculated by convolving in frequency the power exiting the front-end module with the back-end insertion gain. Since the RF insertion gain of these particular devices was unknown we have estimated the magnitude of this approximation by using the insertion gain measured on a different, but similar back-end module. Although non rigorous, this comparison (shown in Figure~\ref{fig:attenuation_power_meter_comparison_insertion_gain}) demonstrates that the power meter measurements provide a good approximation of the back-end module input power.
    
\begin{figure}[h!]
        \begin{center}
            \includegraphics[width=10cm]{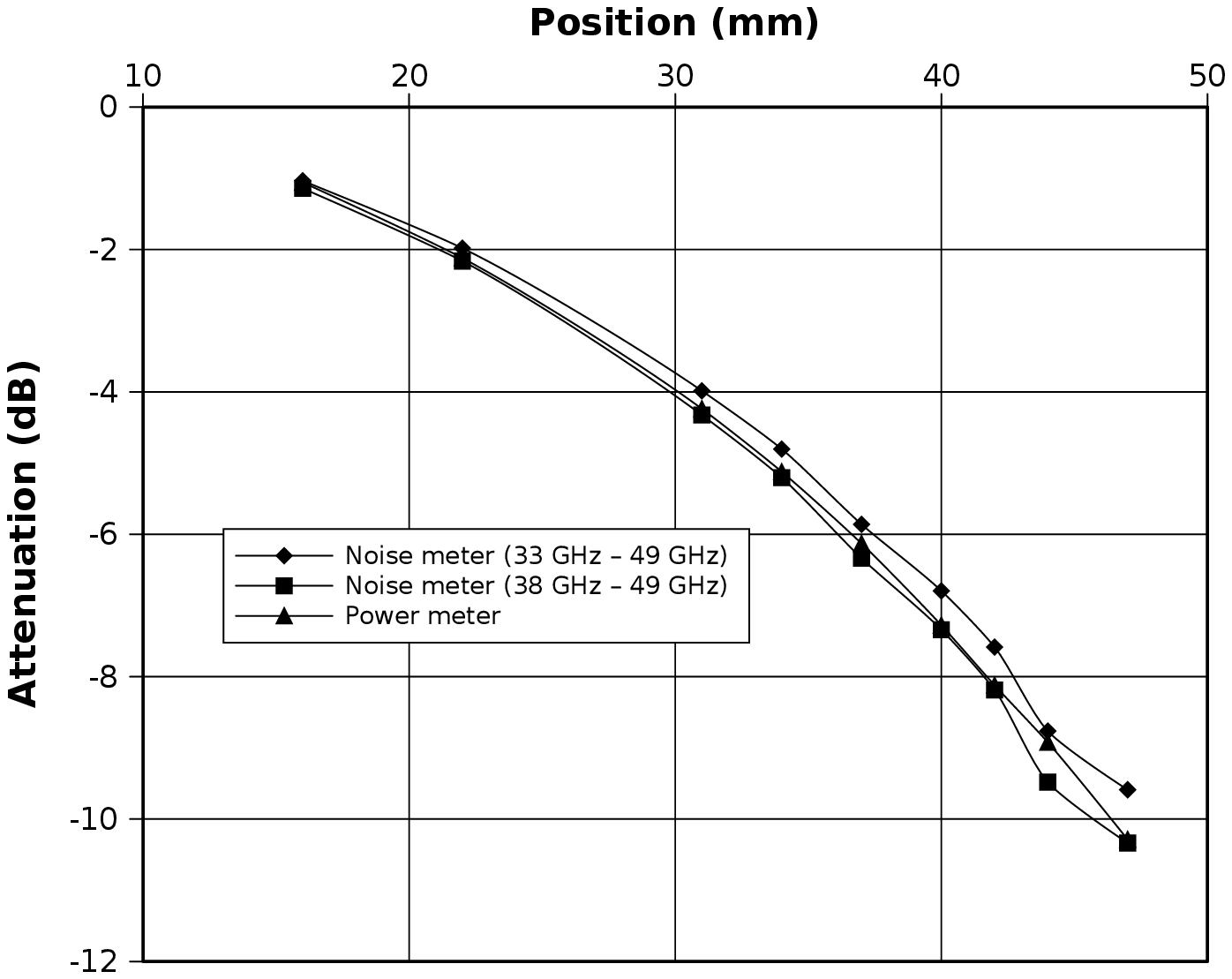}
        \end{center}
        \caption{Comparison between integrated power in two different intervals ([33 GHz -- 49 GHz] and [38 GHz -- 49 GHz]) using the noise meter and power detected using the power meter in the frequency interval [33 GHz -- 50 GHz].}
        \label{fig:attenuation_noise_meter}
    \end{figure}

    \paragraph{Attenuation curves using a noise figure meter.} 
    A noise figure meter was also used to measure power exiting the FEM for several positions of the variable attenuator, roughly corresponding to steps of 1 dB. For each position values have been integrated along the bandwidth and compared with those obtained with the power meter. In Figure~\ref{fig:attenuation_noise_meter} we show the results obtained with the noise figure meter integrated in two different frequency ranges compared with the power meter measurements. The results indicate a good matching of the curves obtained with the different methods.
    
\begin{figure}[h!]
        \begin{center}
            \includegraphics[width=10cm]{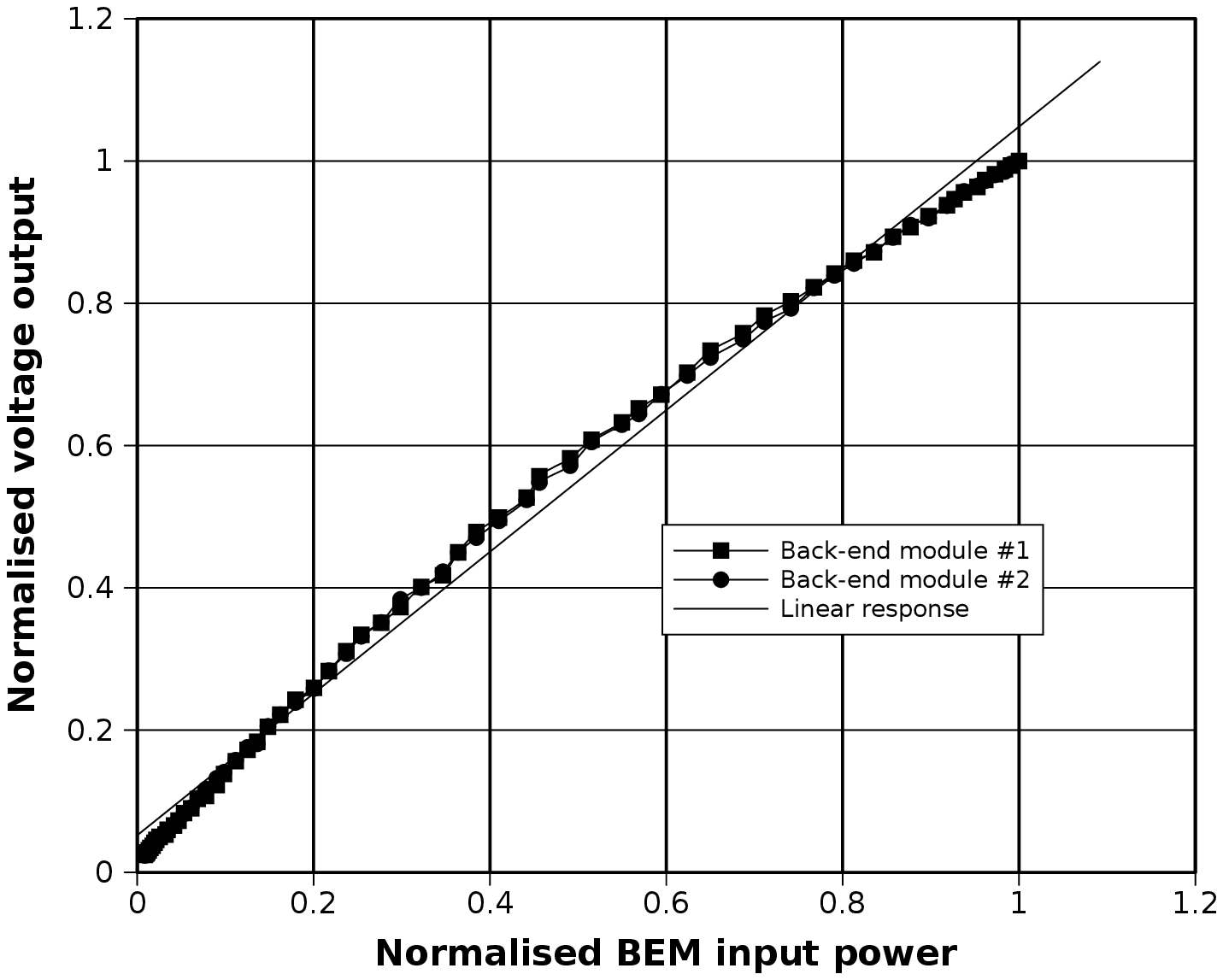}
        \end{center}
        \caption{Normalized compression curves for the two tested 44 GHz back-end modules.}
        \label{fig:normalised_compression_curves}
    \end{figure}

    These results eventually led us to use the average power meter measurements (see Figure~\ref{fig:attenuation_power_meter}) to convert the raw attenuation in mm into power units. In Figure~\ref{fig:normalised_compression_curves} we show the normalised compression curves for the two tested back end modules highlighting the deviation from the expected linear behaviour. Considering that the maximum power corresponded to $\sim 25$~K input temperature, the power range spanned by this test extends well into the temperature region where the receivers will operate in flight, i.e. with few K input temperature.
    
\begin{figure}[h!]
        \begin{center}
            \includegraphics[width=10cm]{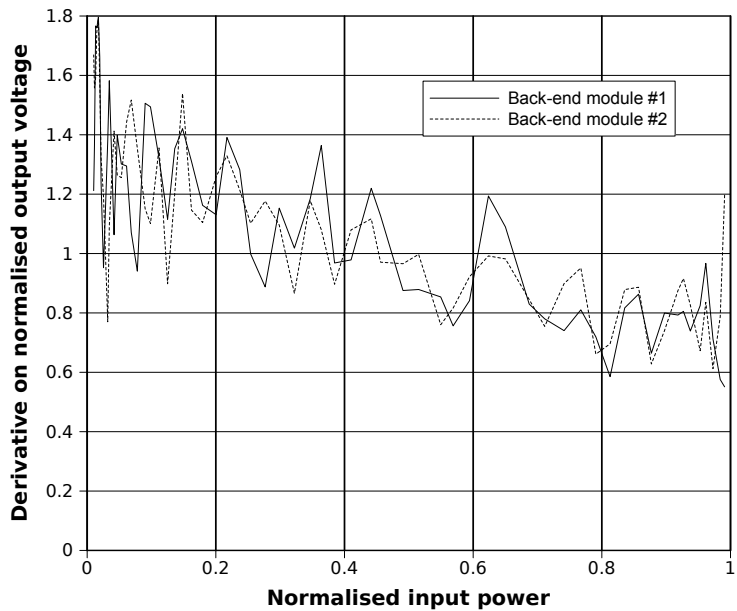}
        \end{center}

        \caption{Derivative of the compression curves. Notice how no region of constant derivative (i.e. linear behaviour) can be found.}
        \label{fig:derivative_normalised_compression_curves}
    \end{figure}

    Analysing the derivative of the compression curves (shown in Figure~\ref{fig:derivative_normalised_compression_curves}) it was apparent that no truly linear response was found across all the input power range.

%

\section{Impact of compression of on ground calibration}
\label{sec:impact_ground_calibration}

  In this section we discuss in detail the effects of compression in the 30 and 44 GHz LFI receivers on ground calibration activities. In particular the following parameters have been calculated with formulas and methods that take into account the non-linear receiver response described by Eq.~(\ref{eq:vout_full}):

\begin{itemize}
    \item noise temperature and photometric calibration constant;
    \item calibrated in-flight white noise sensitivity;
    \item noise effective bandwidth.
\end{itemize}

\subsection{Noise temperature and photometric calibration constant}
\label{sec:noise_t_calibration_const}

    Noise temperature and photometric calibration constant have been calculated from experimental datasets in which the sky-load temperature was varied in a range between $\sim 8$~K and $\sim 30$~K. In the 30 and 44~GHz receivers for each detector we fitted the  $V_{\rm out}(T_{\rm in}^{\rm ant})$ data against Eq.~(\ref{eq:vout_full}) to retrieve $G_0$, $T_{\rm noise}$ and $b$.
    
    In Figure~\ref{fig:non_linear_fit_results} we show an example of the best fit for a 30~GHz and a 44~GHz receiver, while in Appendix~\ref{sec:best_fits} we display the whole set of best fits for the 30~GHz and 44~GHz detectors. The list of the best-fit parameters is reported in Table~\ref{tab:best_fit_parameters}. Further details about tests and data analysis leading to these values can be found in \cite{2009_LFI_cal_M4}.
    
    \begin{figure}[h!] 
        \begin{center}
            \includegraphics[width=7.5cm]{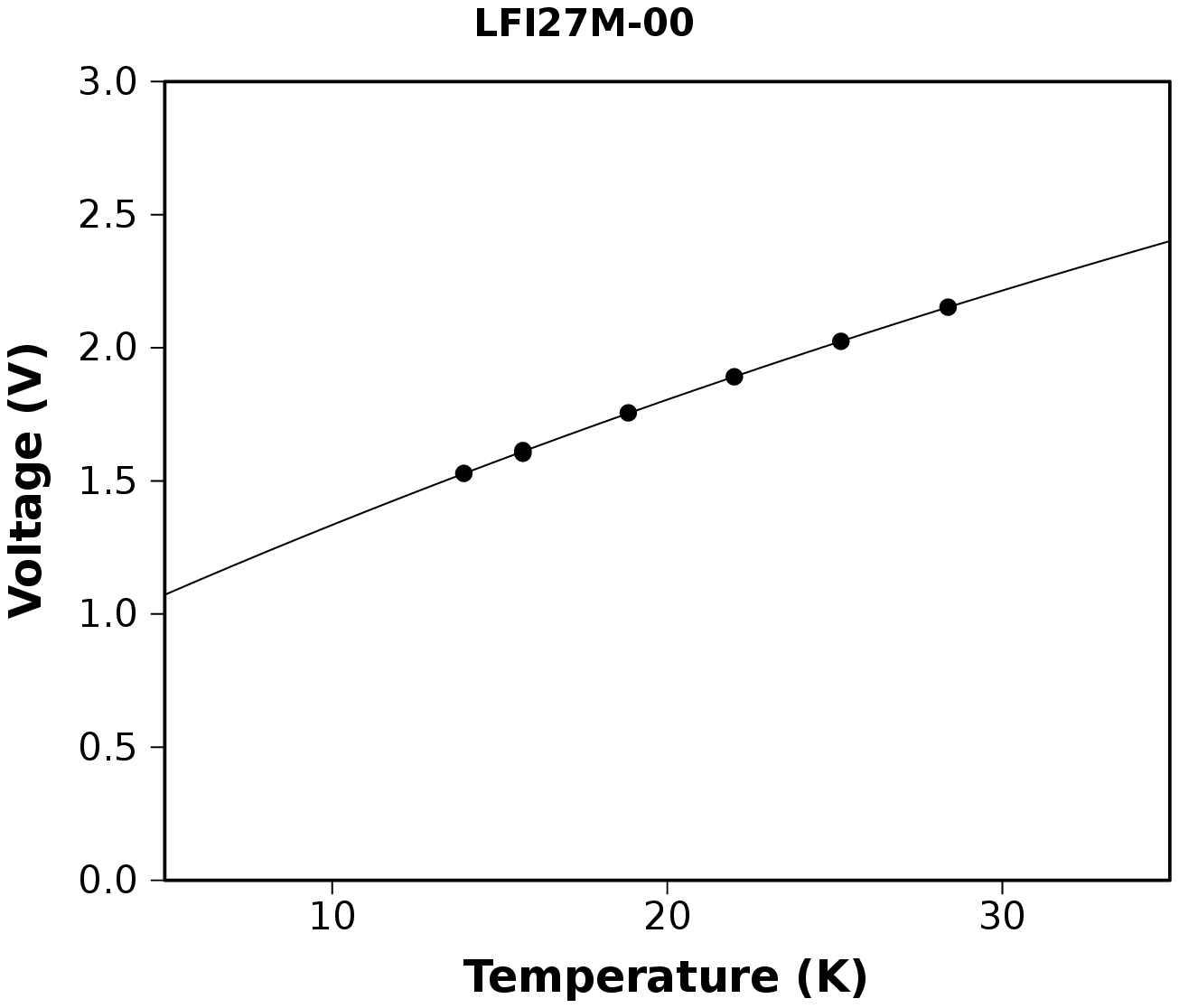}
            \includegraphics[width=7.5cm]{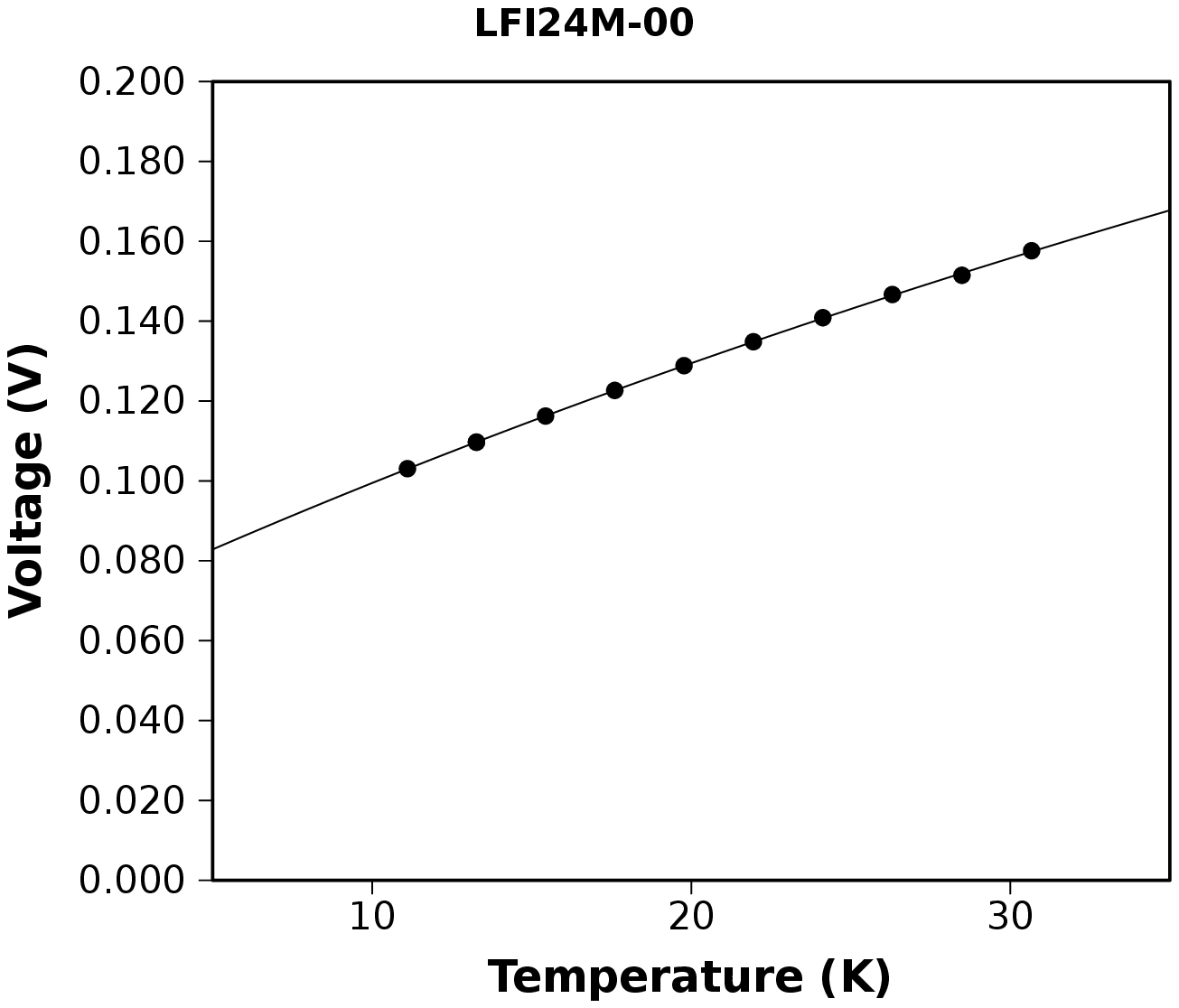}
        \end{center}
        \caption{Two examples of non-linear fitting of $V_{\rm out}$ vs. $T_{\rm in}^{\rm ant}$ data. left panel: 30~GHz receiver LFI27 (detector M-00); right panel: 44~GHz receiver LFI24 (detector M-00).}
        \label{fig:non_linear_fit_results}
    \end{figure}

    \begin{table}[h!]
        \caption{photometric calibration constant, noise temperature and non-linearity parameters obtained from the RCA test campaign (see \cite{2009_LFI_cal_M4}).}
        \label{tab:best_fit_parameters}
        \begin{center}
            \begin{tabular}{|l c c c c|}        
                \multicolumn{5}{c}{$G_0$ (V/K)}   \\
                \hline
                \hline        
                        &M-00   &M-01   &S-10   &S-11\\
                \hline
                LFI24 &  0.0048 & 0.0044 & 0.0062 & 0.0062\\
                LFI25 &  0.0086 & 0.0085 & 0.0079 & 0.0071\\
                LFI26 &  0.0052 & 0.0067 & 0.0075 & 0.0082\\
                LFI27 &  0.0723 & 0.0774 & 0.0663 & 0.0562\\
                LFI28 &  0.0621 & 0.0839 & 0.0607 & 0.0518\\
            \hline
            \end{tabular}
            
            \vspace{1cm}
            \begin{tabular}{|l c c c c|}
                \multicolumn{5}{c}{$T_{\rm noise}$ (K)}   \\
                \hline
                \hline        
                        &M-00   &M-01   &S-10   &S-11\\
                \hline
                LFI24   &15.5  &  15.3  &  15.8  &  15.8\\
                LFI25   &17.5  &  17.9  &  18.6  &  18.4\\
                LFI26   &18.4  &  17.4  &  16.8  &  16.5\\
                LFI27   &12.1  &  11.9  &  13.0  &  12.5\\
                LFI28   &10.6  &  10.3  &   9.9  &   9.8\\
                \hline   
            \end{tabular}
            \begin{tabular}{|l c c c c|}        
                \multicolumn{5}{c}{$b$}\\
                \hline
                \hline        
                        &M-00   &M-01   &S-10   &S-11\\
                \hline
                    LFI24   &1.794 &  1.486 &  1.444 &  1.446\\
                    LFI25   &1.221 &  1.171 &  0.800 &  1.013\\
                    LFI26   &1.085 &  1.418 &  0.943 &  1.218\\
                    LFI27   &0.123 &  0.122 &  0.127 &  0.140\\
                    LFI28   &0.190 &  0.157 &  0.187 &  0.196\\
            \hline
            \end{tabular}
        \end{center}
    \end{table}

\subsection{Calibrated in-flight sensitivity}
\label{sec:calibrated_in-flight_sensitivity}

    One of the key performance parameters derived from data acquired during the calibration campaign is the in-flight calibrated sensitivity, estimated starting from the raw uncalibrated white noise sensitivity measured at laboratory conditions which were similar but not equal to the expected in flight conditions. In particular during laboratory experiments the input sky temperature was $\gtrsim 8$~K and the front-end unit temperature was, in some cases (e.g. during the instrument-level test campaign~\cite{2009_LFI_cal_M3}) greater than 20~K. 

    In this section we discuss how the raw noise measurements in the laboratory have been extrapolated to flight conditions with particular reference to the effect of response non-linearity on the calculations.
    
    Our starting point is the the raw datum, a couple of uncalibrated white noise levels in V$\times\sqrt{\rm s}$ for the two detectors in a radiometer measured with the sky load at a temperature $T_{\rm {sky-load}}$ and the front end unit at physical temperature $T_{\rm test}$. 
    In order to derive the calibrated white noise level extrapolated to input temperature equal to $T_{\rm sky}$ and with the front end unit at a temperature of $T_{\rm nominal}$ we have performed the following three steps:
    
    \begin{enumerate}
        \item extrapolation to nominal front-end unit temperature;
        \item extrapolation to nominal input sky temperature;
        \item calibration in units of  K$\times \sqrt{\rm s}$.
    \end{enumerate}

    A detailed discussion of the first step can be found in \cite{2009_LFI_cal_M3}. Here we will focus on the other points, which are affected by non linearity in the receiver response.
        
            
    Let us start from the radiometer equation in which, for each detector, the white noise spectral density is given by:
    
    \begin{equation}
        \delta T_{\rm rms} = 2\frac{T_{\rm in}+T_{\rm noise}}{\sqrt{\beta}}
        \label{eq:single_diode_radiometer_equation}
    \end{equation}
    
    Now we want to find a similar relationship for the uncalibrated white noise spectral density linking $\delta V_{\rm rms}$ to $V_{\rm out}$. We start from the following:
    
    \begin{equation}
        \delta V_{\rm rms} = \frac{\partial V_{\rm out}}{\partial T_{\rm in}}\delta T_{\rm rms};
        \label{eq:delta_vrms_basic}
    \end{equation}
    calculating the derivative of $V_{\rm out}$ using Eq.~(\ref{eq:vout_full}) and using $\delta T_{\rm rms}$ from Eq.~(\ref{eq:single_diode_radiometer_equation}) we obtain:
    
    \begin{equation}
        \delta V_{\rm rms} = \frac{V_{\text{out}}}{\sqrt{\beta }}\left[1+ 
        G_0 b \left(T_{\text{in}}+T_n\right)\right]^{-1},
        \label{eq:white_noise}
    \end{equation}
    
    where $\beta$ is the bandwidth and $V_{\rm out}$ is the receiver DC voltage output. Considering the two input temperatures $T_{\rm sky-load}$ and $T_{\rm sky}$ then the ratio  $\rho = \frac{\delta V_{\rm rms}(T_{\rm sky})}{\delta V_{\rm rms}(T_{\rm sky-load})}$ is:
    
    \begin{equation}
        \rho = \frac{ V_{\rm out}(T_{\rm sky})}
        {V_{\rm out}(T_{\rm sky-load})}
        \frac{1+G_0 b(T_{\rm sky-load}+T_{\rm noise})}{1+G_0 b(T_{\rm sky}+T_{\rm noise})}.
        \label{eq:ratio_uncalibrated_white_noise}
    \end{equation}
    
     Using Eq.~(\ref{eq:vout_full}) to expand $\rho$ in Eq.~(\ref{eq:ratio_uncalibrated_white_noise}) we have:
    
     \begin{equation}
        \rho = \frac{
        T_{\rm sky}+T_{\rm noise}}{T_{\rm sky-load}+T_{\rm noise} }
        \left[\frac{1+ b\, G_0
        (T_{\rm sky-load}+T_{\rm noise})}{1+b\, G_0
        (T_{\rm sky}+T_{\rm noise})}\right]^2,
        \label{eq:ratio_uncalibrated_white_noise_1}
    \end{equation}
    and $\delta V_{\rm rms}(T_{\rm sky}) = \rho\times \delta V_{\rm rms}(T_{\rm chamber})$.
    
    From Eqs.~(\ref{eq:white_noise}) and (\ref{eq:vout_full}) we obtain that
    
    \begin{equation}
        \delta V_{\rm rms} = \frac{G_0}{\left[1+b\, G_0(T_{\rm sky}+T_{\rm noise})\right]^2}\times 2\frac{T_{\rm sky}+T_{\rm noise}}{\sqrt{\beta}}.
        \label{eq:tilde_wn_final}
    \end{equation}
    
    The calibrated noise extrapolated at the sky temperature, $\delta T_{\rm rms}$, can be obtained considering that, by definition, $\delta T_{\rm rms} = 2\frac{T_{\rm sky}+T_{\rm noise}}{\sqrt{\beta}}$, therefore:
    
    \begin{equation}
        \delta T_{\rm rms} =  \frac{\left[1+b\, G_0(T_{\rm sky}+T_{\rm noise})\right]^2}{G_0} \delta V_{\rm rms}.
    \end{equation}

    A summary of the expected in-flight sensitivities for the Planck-LFI can be found in \cite{2009_LFI_cal_M3}.

\subsection{Noise effective bandwidth}
\label{sec_noise_effective_bandwidth}

   The well-known radiometer equation applied to the single-diode output links the white noise level to sky and noise temperatures and the receiver bandwidth. It reads \cite{seiffert02}:

   \begin{equation}
      \delta T_{\rm rms} = 2\frac{T_{\rm sky}+T_{\rm noise}}{\sqrt{\beta}}.
      \label{eq:radiometer_equation}
   \end{equation}

   In the case of linear response we can write Eq.~\ref{eq:radiometer_equation} in its most useful uncalibrated form:

   \begin{equation}
      \delta V_{\rm rms} = 2\frac{V_{\rm out}}{\sqrt{\beta}},
      \label{eq:radiometer_equation_uncalibrated}      
   \end{equation}
   which is commonly used to estimate the receiver bandwidth, $\beta$, from a simple measurement of the receiver DC output and white noise level, i.e.:

   \begin{equation}
      \tilde\beta = 4\left(\frac{V_{\rm out}}{\delta V_{\rm rms}}\right)^2.
      \label{eq:noise_effective_bandwidth}
   \end{equation}

   If the response is linear and if the noise is purely radiometric (i.e. all the additive noise from back end electronics is negligible and if there are no non-thermal noise inputs from the source) then $\tilde \beta$ is equivalent to the receiver bandwidth, i.e. 

   \begin{equation}
      \tilde \beta \equiv \beta = 4\left(\frac{T_{\rm sky}+T_{\rm noise}}{\delta T_{\rm rms}}\right)^2.
      \label{eq:bandwidths_equivalence}
   \end{equation}

   Conversely, if the receiver output is compressed, from Eq.~(\ref{eq:vout_full}) we have that:
   
   \begin{equation}
      \delta V_{\rm rms} = \frac{\partial V_{\rm out}}{\partial T_{\rm in}}\delta T_{\rm rms}.
      \label{eq:delta_vrms}
   \end{equation}

\begin{figure}[h!]
        \begin{center}
            \includegraphics[width=7.5cm]{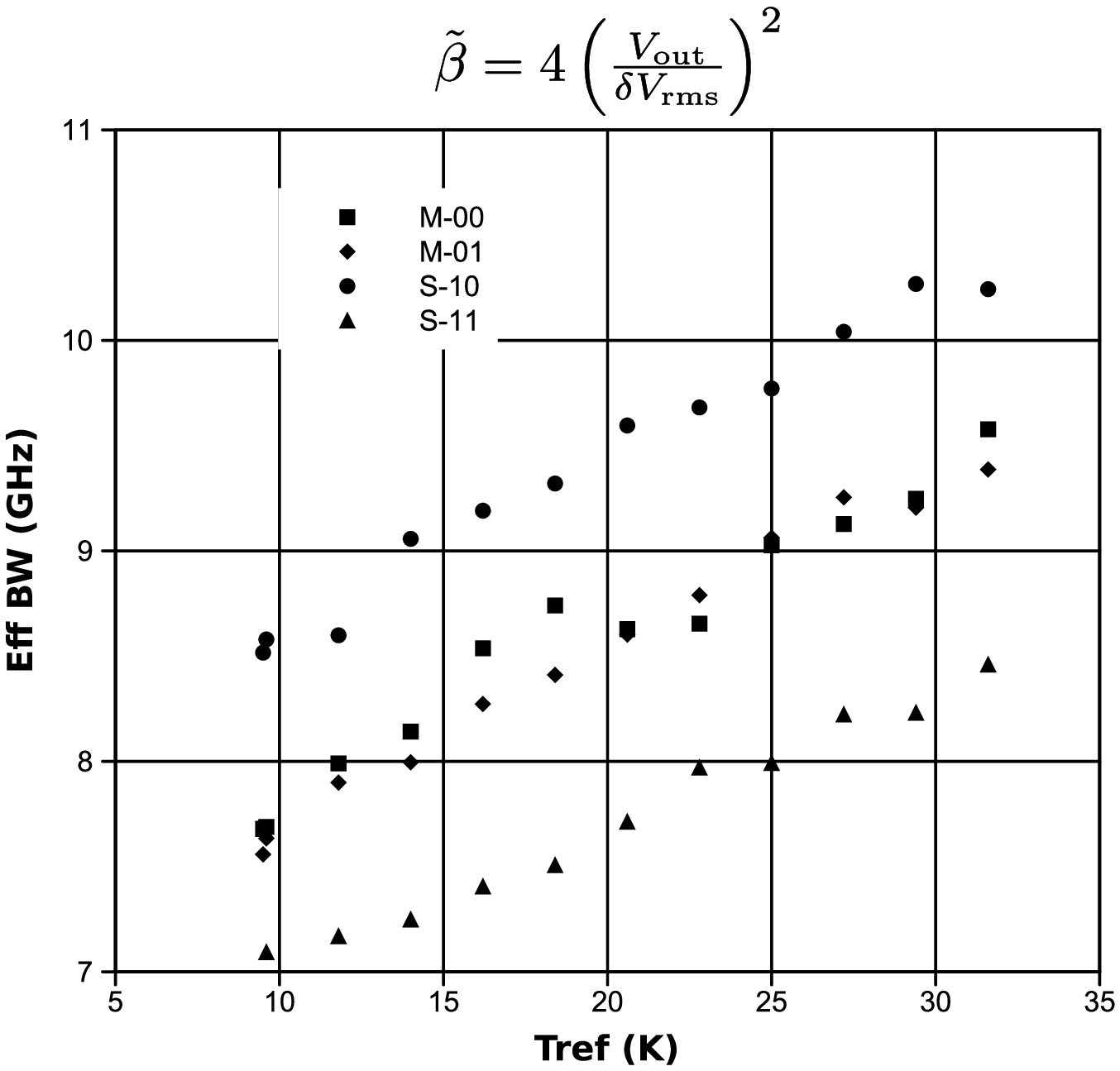}
            \includegraphics[width=7.5cm]{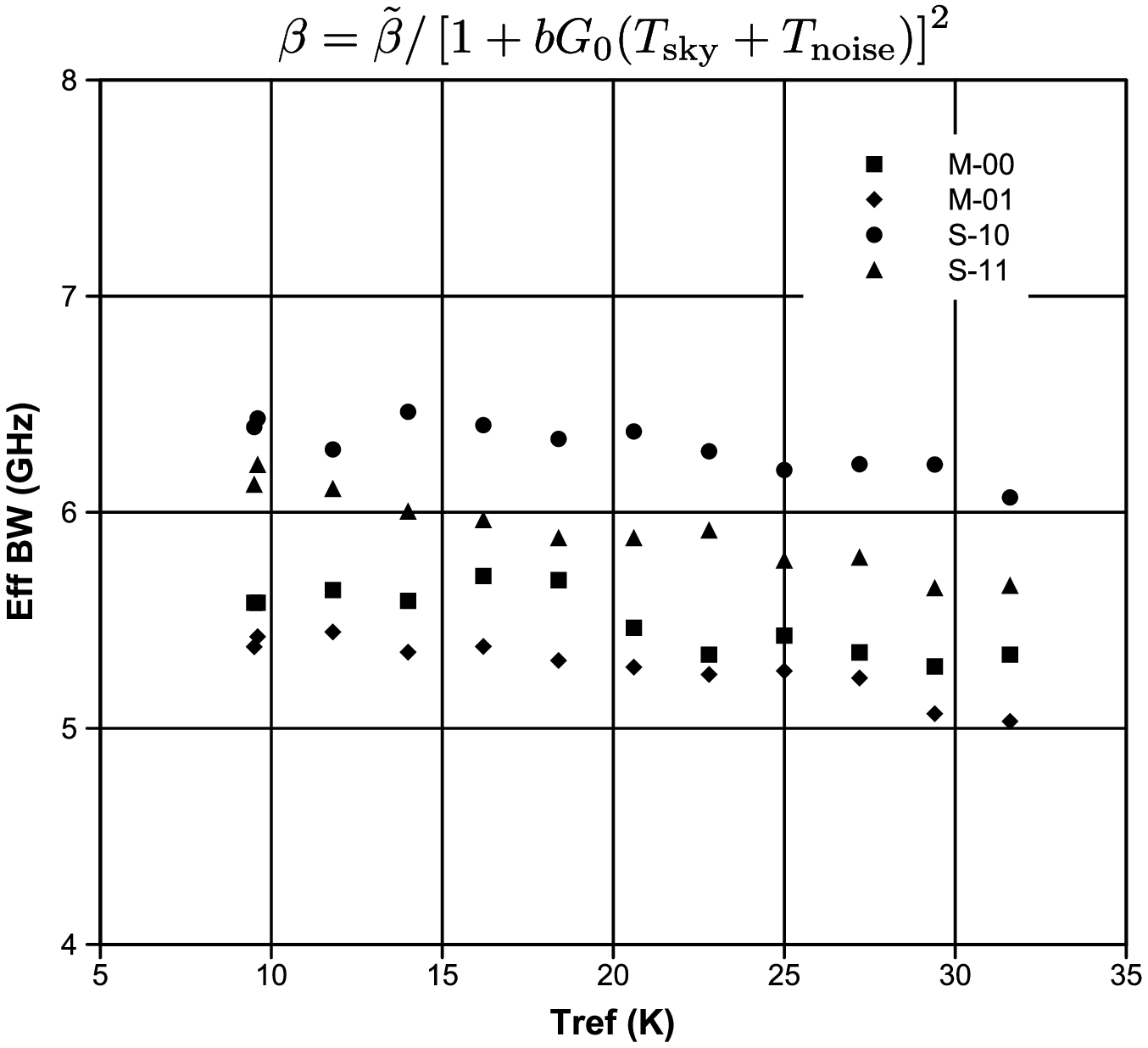}
        \end{center}
        \caption{Noise effective bandwidth 30~GHz receiver LFI27 calculated with different reference load input temperatures
        neglecting (left) and considering (right) the compresson effect.}
        \label{fig:eff_bw_lfi27}
   \end{figure}

   By combining Eqs.~ (\ref{eq:vout_full}), (\ref{eq:noise_effective_bandwidth}) and  (\ref{eq:delta_vrms}) we find:
   
   \begin{equation}
      \tilde \beta = 4\left(\frac{T_{\rm sky}+T_{\rm noise}}{\delta T_{\rm rms}}\right)^2 
     \left[ 1 + b\, G_0(T_{\rm sky}+T_{\rm noise})\right]^2 \equiv
      \beta  \left[ 1 + b\, G_0(T_{\rm sky}+T_{\rm noise})\right]^2,
      \label{eq:bandwidth_compressed}
   \end{equation}
   which shows that $\tilde \beta$ overestimates the ``optical'' bandwidth unless the non linearity parameter $b$ is very small. In the left panel of Figure~\ref{fig:eff_bw_lfi27} we show how the noise effective bandwidth calculated from Eq.~(\ref{eq:noise_effective_bandwidth}) is dependent from the level of the input signal if the receiver response is non linear. In the right panel of the same figure we show how the dependence on the level of the input signal disappears if we take into account the receiver non linearity via Eq.~(\ref{eq:bandwidth_compressed}).
   
   Data presented in Figure~\ref{fig:eff_bw_lfi27} have been taken during the RCA test campaign with various levels of the reference load temperature and refer to the 30 GHz receiver LFI27.
   
\begin{figure}[h!]
        \begin{center}
            \includegraphics[width=9cm]{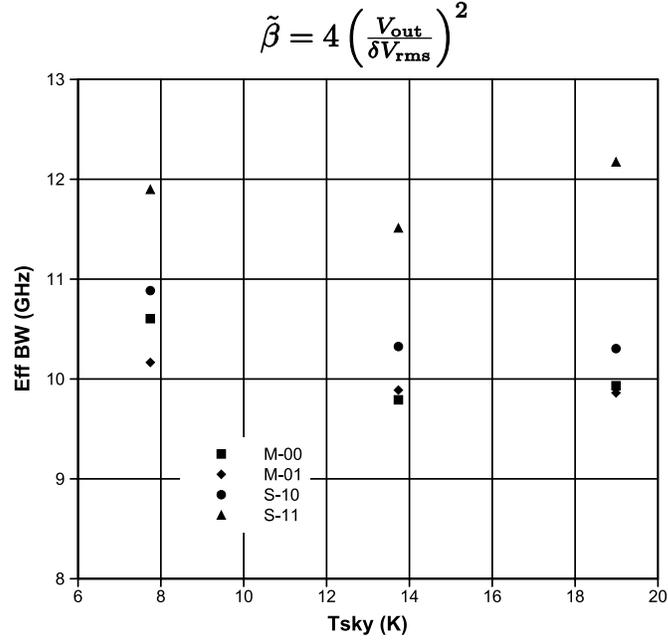}
        \end{center}
        \caption{Noise effective bandwidth for the 70 GHz receiver LFI19 with different sky load input temperatures calculated without compression effect. 
        }
        \label{fig:eff_bw_lfi19}
   \end{figure}

    In Figure~\ref{fig:eff_bw_lfi19} we show similar data for the 70~GHz receiver LFI19. Data were acquired during the RCA test campaign with a variable input temperature at the sky load. In this case data clearly do not show a consistent trend in the noise effective bandwidth calculated from Eq.~(\ref{eq:noise_effective_bandwidth}) with input temperature, which provides an independent confirmation of the linear response of the 70~GHz receivers.
    
\section{Impact of output compression in flight operations}
\label{sec:impact_flight_operations}

  In this section we analyse the typical signal dynamic range that will be encountered in flight to verify whether in these conditions the receiver response can be considered linear or not.
This is important especially for the in-flight calibration activities, in particular:
\begin{itemize}
    \item for photometric calibration, performed continuously by exploiting the well-known dipole anisotropy \cite{cappellini03}, and
    \item for main beam measurements, carried out thanks to bright point sources like Jupiter and Saturn \cite{burigana01a}.
\end{itemize}

Let us consider the LFI observing the sky with a temperature $T_{\rm sky}+\delta T$ with a stable reference load temperature $T_{\rm ref}$. The uncalibrated differential output from a single diode is:

\begin{equation}
    \delta V = V_{\rm sky} - r\times V_{\rm ref},
    \label{eq:differential_output}
\end{equation}
where $r = \frac{T_{\rm sky}+T_{\rm noise}}{T_{\rm sky}+T_{\rm noise}}$. Expanding $V_{\rm sky}$ and $V_{\rm ref}$ using Eq.~(\ref{eq:vout_full}) we obtain $\delta V = G \, \delta T$ where:

\begin{eqnarray}
     G = G_0\left\{\left[1  \right. \right.&+&b\, G_0 (T_{\rm sky}+ \left.T_{\rm noise})\right] \times\nonumber\\ 
     &\times&\left.\left[1+b\, G_0 (T_{\rm sky}+\delta T + T_{\rm noise})\right]\right\}^{-1}
     \label{eq:in_flight_photometric_calibration}
\end{eqnarray}

The relative variation in the photometric calibration constant, $\delta G / G$, caused by a variation $\delta T$ in the input temperature can be calculated by:

\begin{equation}
    \frac{\delta G}{G} = \frac{1}{G} \frac{\partial G}{\partial (\delta T)} \delta T = 
    - \frac{b\, \delta T \, G_0}{1 + b\, G_0 ( T_{\rm sky} + \delta T + T_{\rm noise})}
\end{equation}

If we now estimate $\delta G / G$ assuming $\delta T \sim \pm 3$~mK (dipole anisotropy) and $\delta T\sim \pm 50$~mK (Jupiter) and using the receiver parameters $G_0$, $T_{\rm noise}$ and $b$ listed in Table.~\ref{tab:best_fit_parameters} we find:

\begin{equation}
\begin{array}{l l l}
    \frac{\delta G}{G} \lesssim 6\times 10^{-5} &\mbox{ for } &\delta T\sim \pm 3\, \mbox{mK}\\
    \mbox{}\\
    \frac{\delta G}{G} \lesssim 10^{-3} &\mbox{ for } &\delta T\sim\pm 50\, \mbox{mK},
\end{array}
\end{equation}
which clearly shows how the receiver output can be considered linear with the input signal dynamic range expected during flight nominal operations.

Although the knowledge of the non linear response is not necessary for data analysis of nominal flight data, all the parameters listed in Table~\ref{tab:best_fit_parameters} will be measured during the in-flight calibration and verification phase that will be performed before the start of the nominal operations. In particular the cooldown of the HFI 4~K cooler will provide an input signal varying from $\sim$20~K to the stable nominal temperature of $\sim$4~K, allowing the determination of the linearity parameter.

\section{Conclusions}
\label{sec:conclusions}

  In this paper we have discussed the linearity properties of the Planck-LFI receivers. The voltage output has been measured during the calibration campaign of the individual receivers and of the integrated instrument using a signal input ranging from $\sim$8~K to $\sim$30~K.

The receiver response is linear for the 70~GHz receivers while the 30 and 44~GHz radiometers show slightly compressed response over all the input signal range, which is well described by the relationship in Eq.~(\ref{eq:vout_full}). The source of signal compression has been identified in the back-end amplifiers and detector diodes and characterised by dedicated tests on two 44~GHz back-end modules.

The calculation of several performance parameters from data acquired during ground tests must take into account signal compression in order to provide a correct estimate. In particular the calculation of noise temperature, photometric calibration, white noise sensitivity and noise effective bandwidth can be completely off the correct estimates if nonlinearity is not properly taken into account.

Although compression impacts the calculation of receiver performance parameters from ground test data, it is essentially negligible during nominal operations, where the input signal dynamic range is small enough ($\lesssim\pm 50$~mK) to keep the response in the linear regime. 

The last characterisation of the LFI receivers linearity will be performed during the in flight calibration exploiting the cooldown of the HFI 4~K cooler which will provide an input signal over a range from $\sim 20$~K to $\sim 4$~K.

\acknowledgments
    Planck is a project of the European Space Agency with instruments funded by ESA member states, and with special contributions from Denmark and NASA (USA). The Planck-LFI project is developed by an Interntional Consortium lead by Italy and involving Canada, Finland, Germany, Norway, Spain, Switzerland, UK, USA. The Italian contribution to Planck is supported  by the Italian Space Agency (ASI). The work in this paper has been supported by in the framework of the ASI-E2 phase of the Planck contract. The US Planck Project is supported by the NASA Science Mission Directorate. In Finland, the Planck LFI 70 GHz work was supported by the Finnish Funding Agency for Technology and Innovation (Tekes).
\clearpage
\appendix

    \section{Best fits}
\label{sec:best_fits}
 
\begin{figure}[h!]
    \begin{center}
        \includegraphics[width=3.5cm]{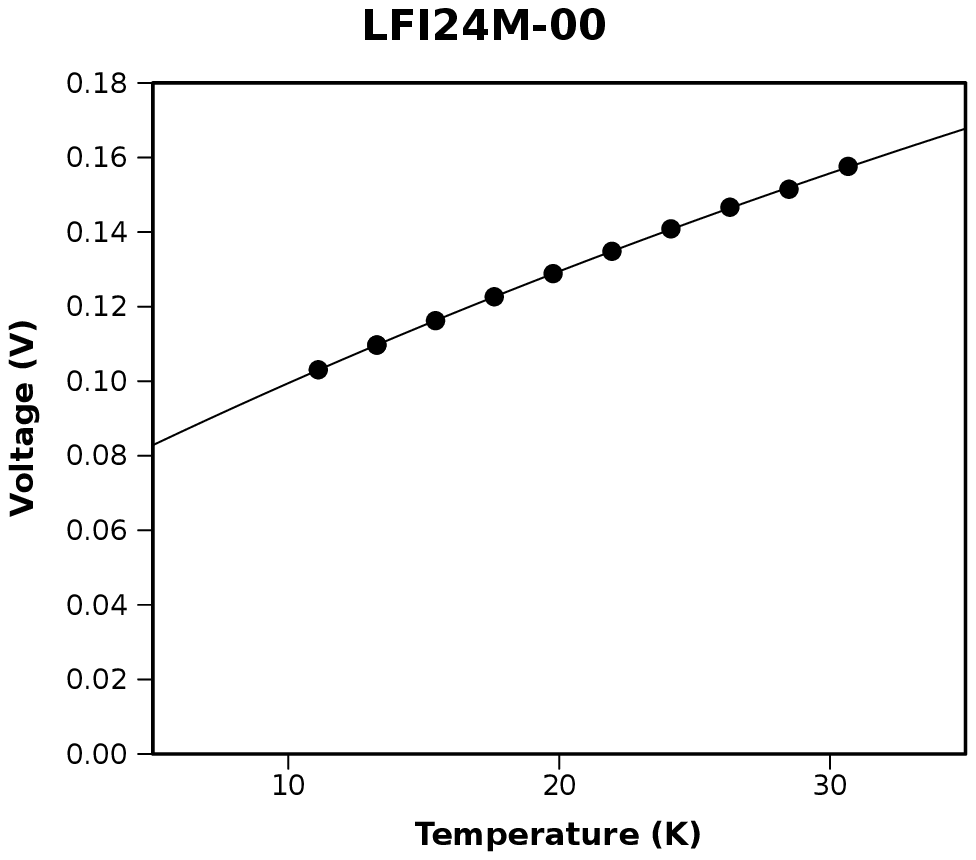}
        \includegraphics[width=3.5cm]{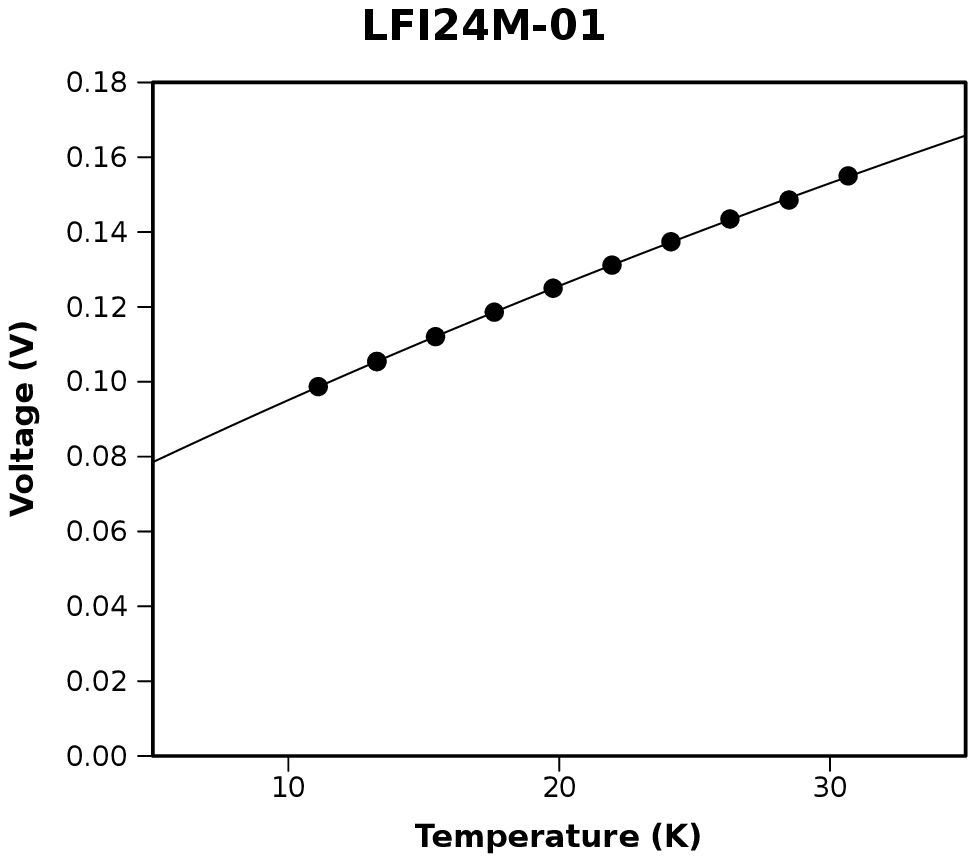}
        \includegraphics[width=3.5cm]{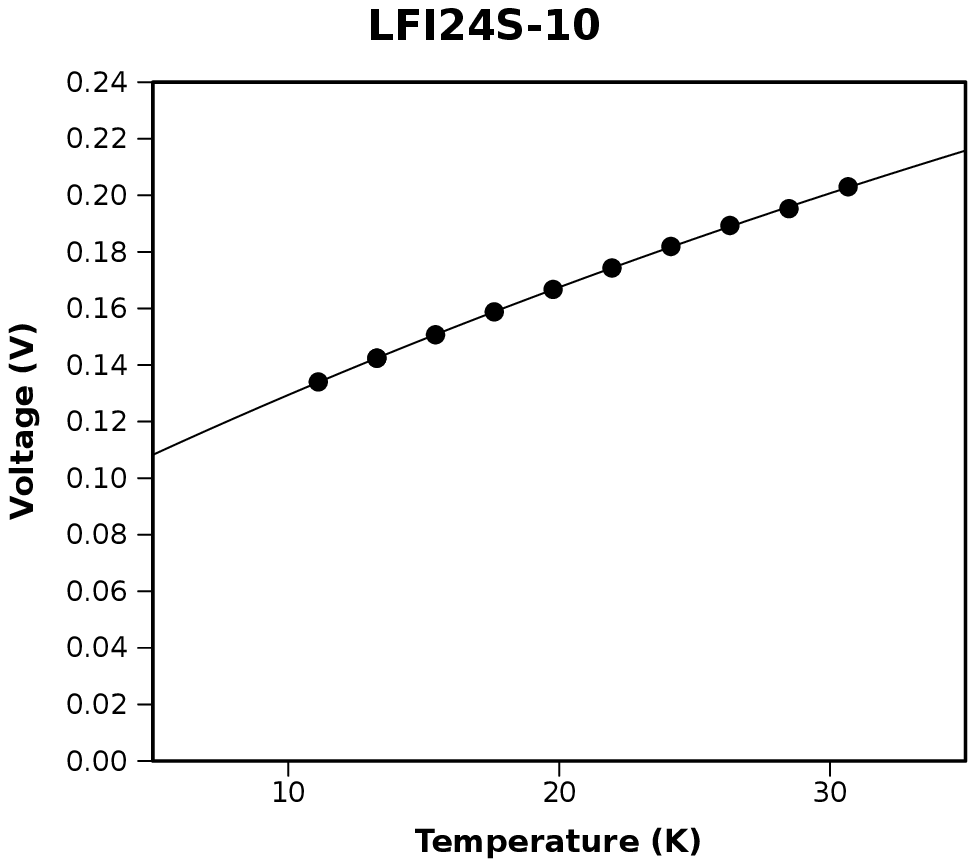}
        \includegraphics[width=3.5cm]{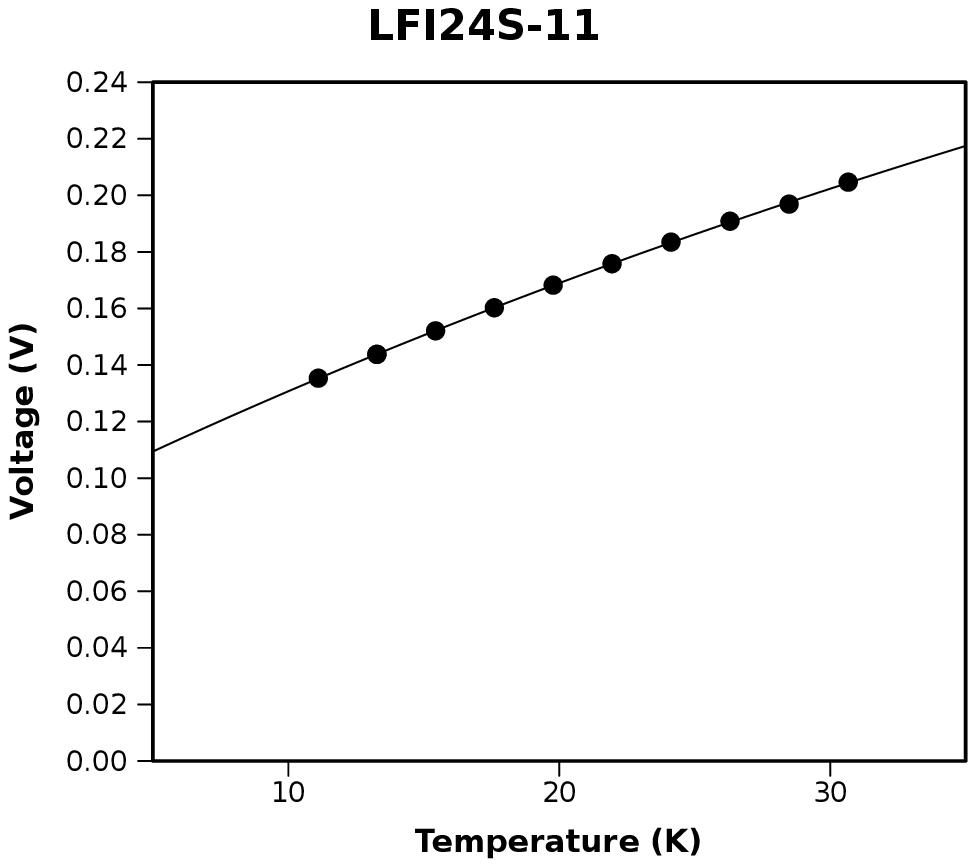}\\
        
        \includegraphics[width=3.5cm]{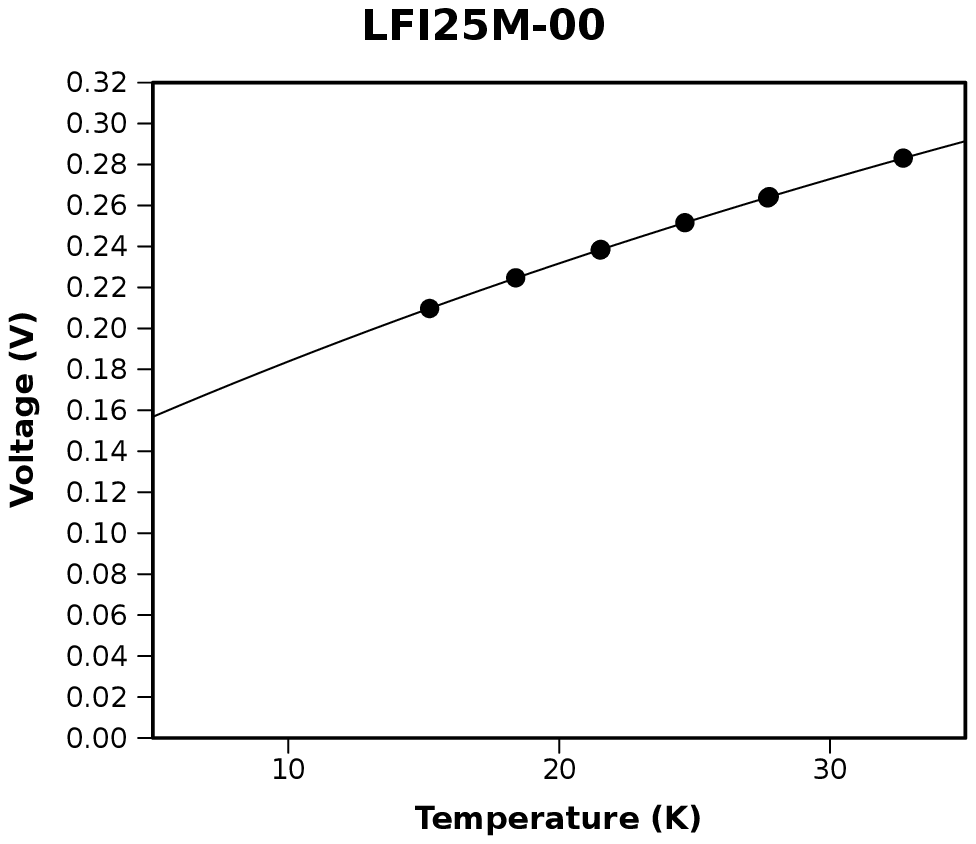}
        \includegraphics[width=3.5cm]{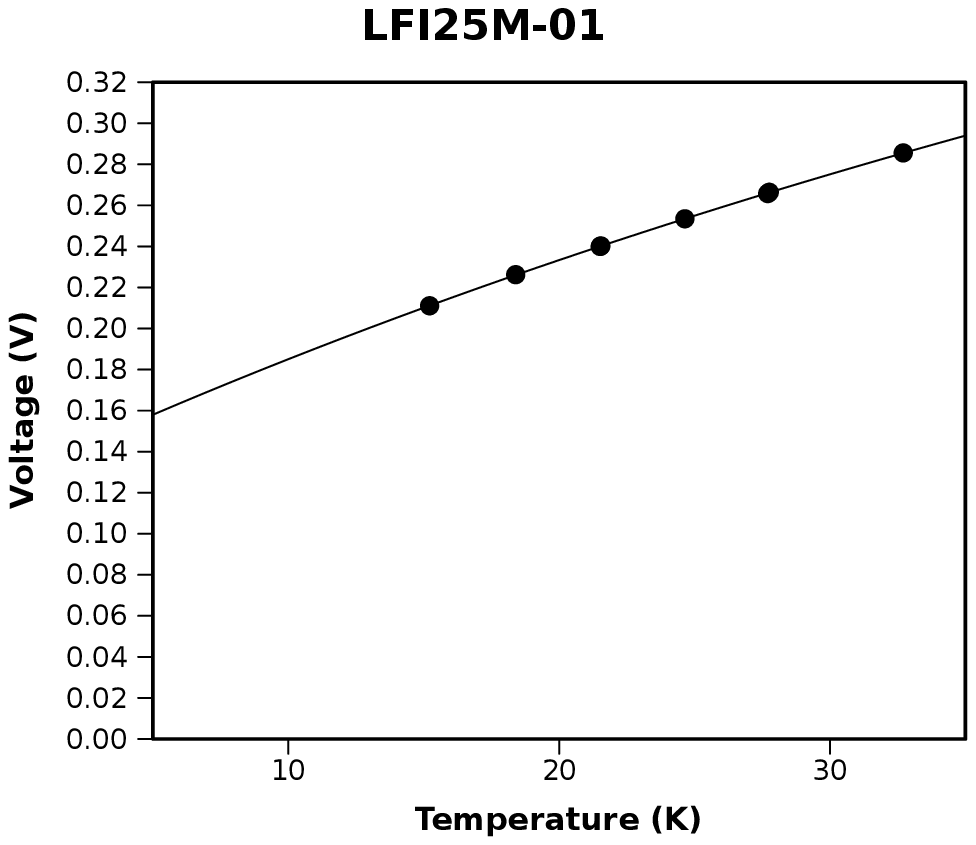}
        \includegraphics[width=3.5cm]{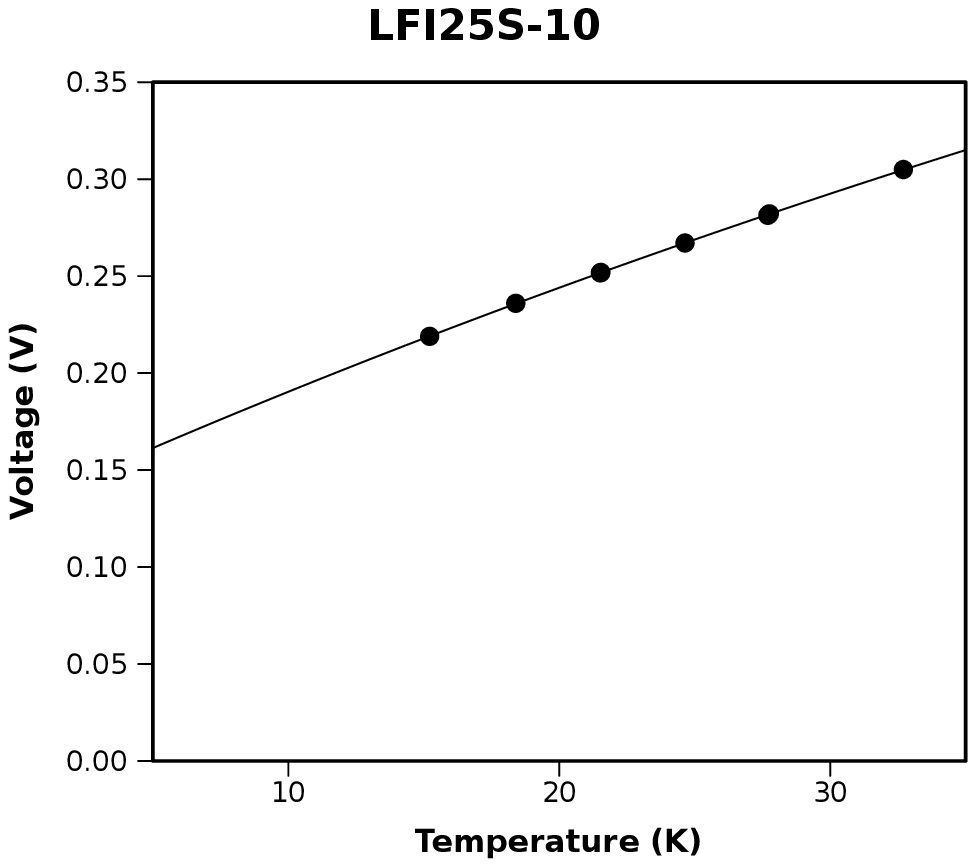}
        \includegraphics[width=3.5cm]{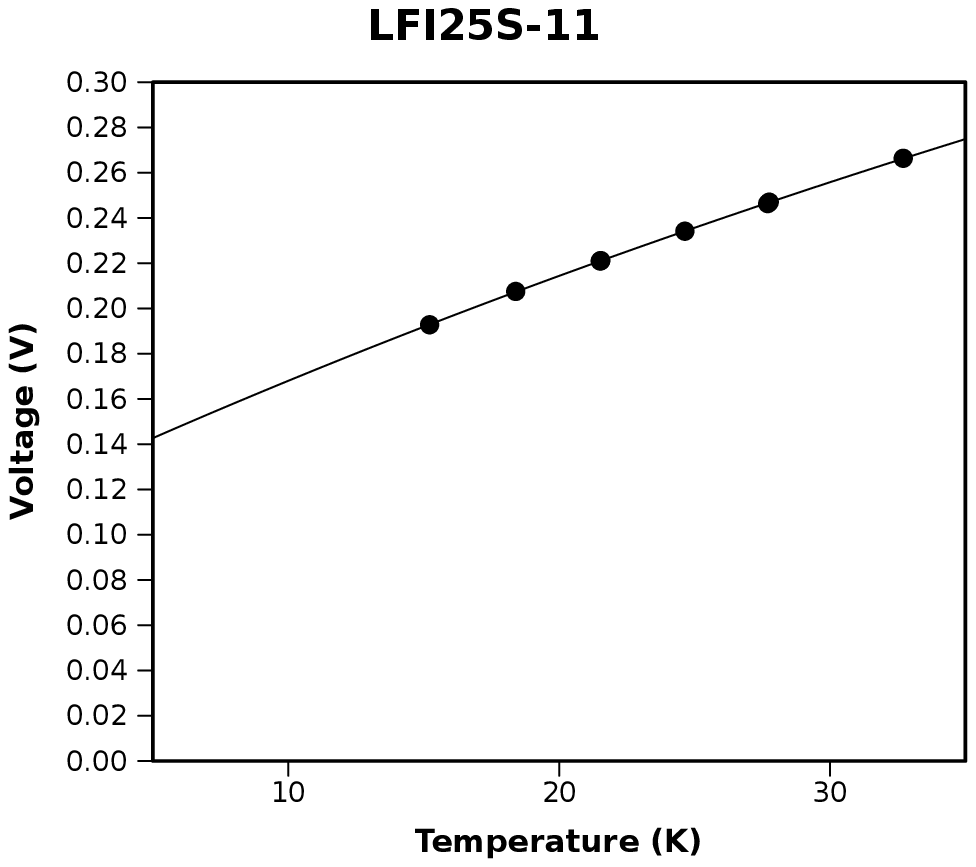}\\
    
        \includegraphics[width=3.5cm]{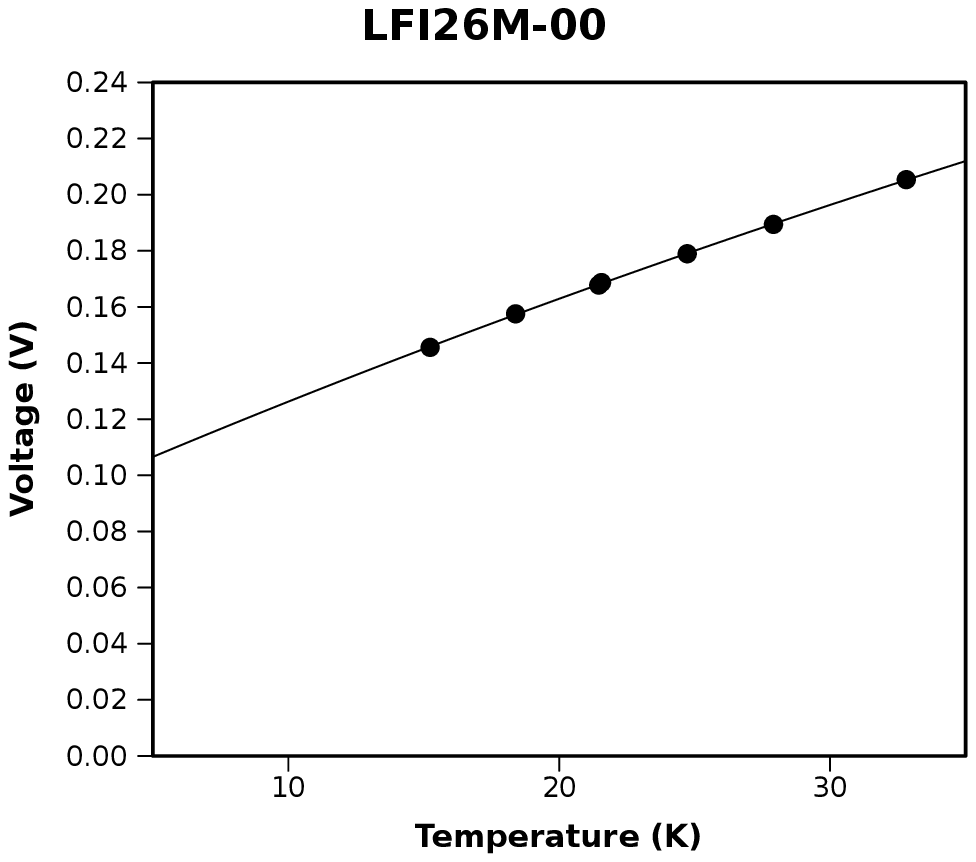}
        \includegraphics[width=3.5cm]{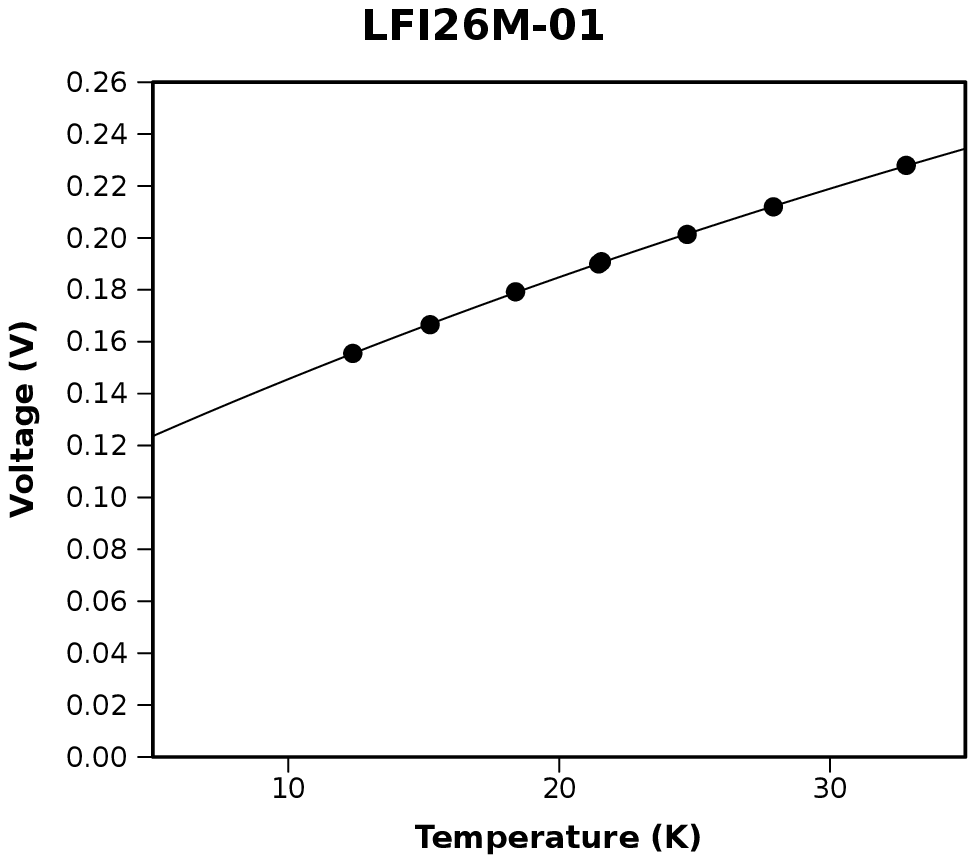}
        \includegraphics[width=3.5cm]{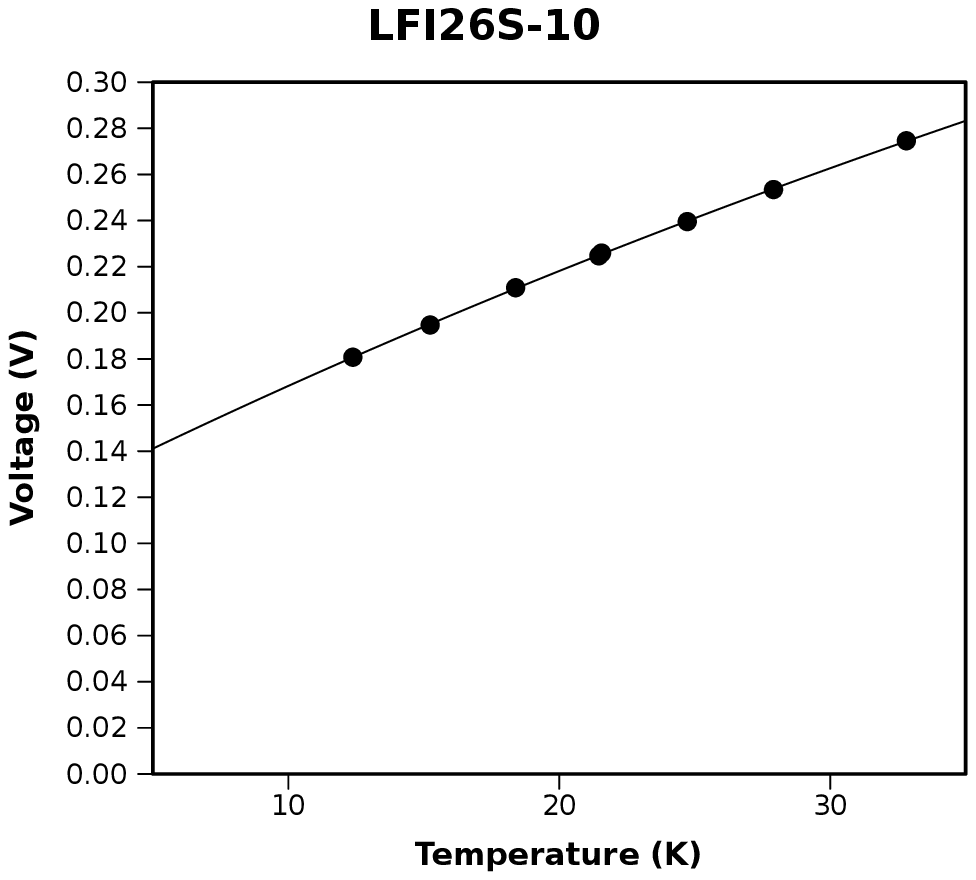}
        \includegraphics[width=3.5cm]{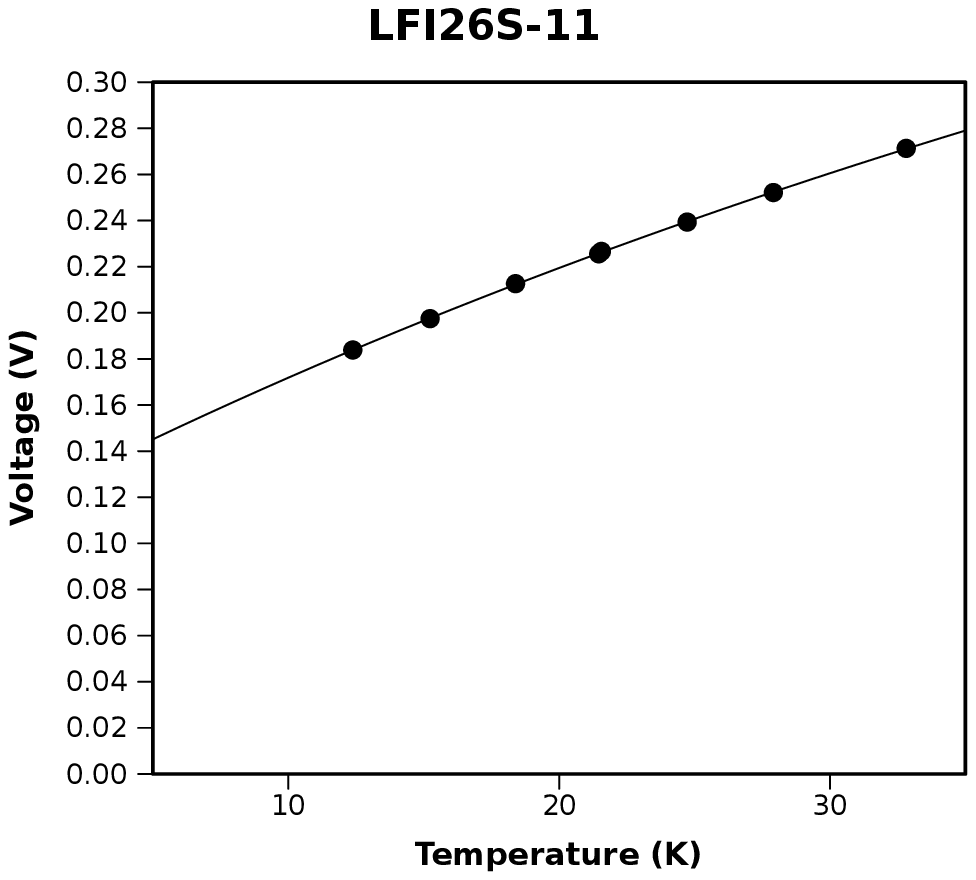}\\

        \includegraphics[width=3.5cm]{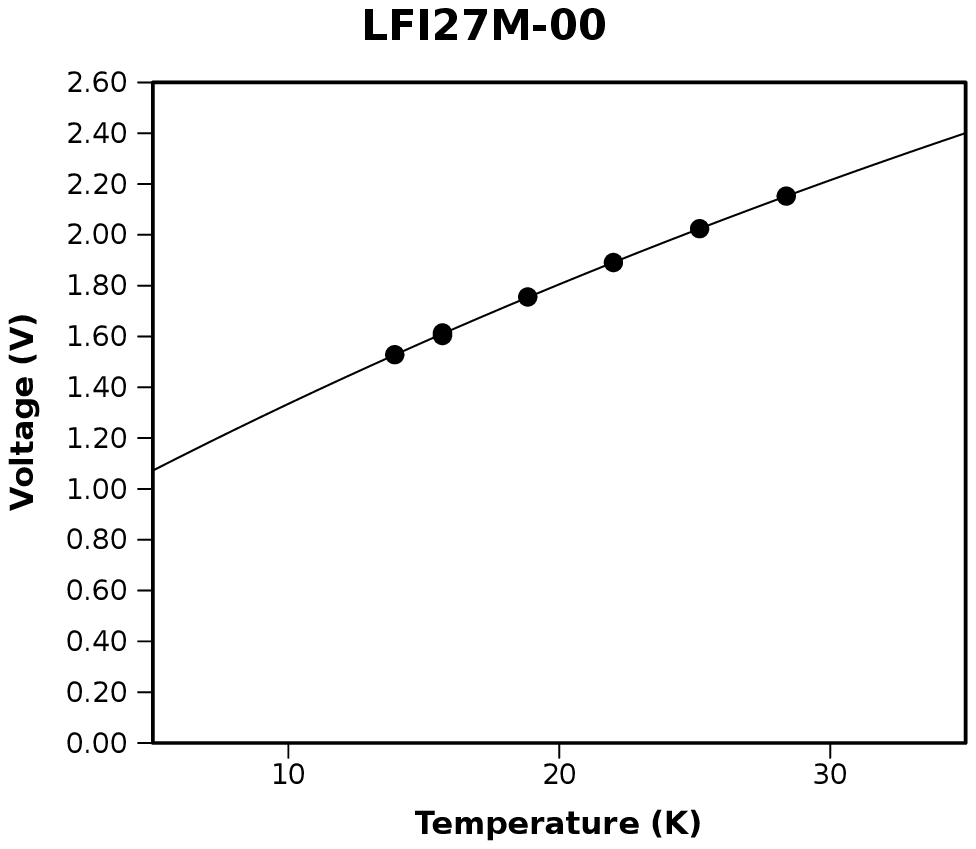}
        \includegraphics[width=3.5cm]{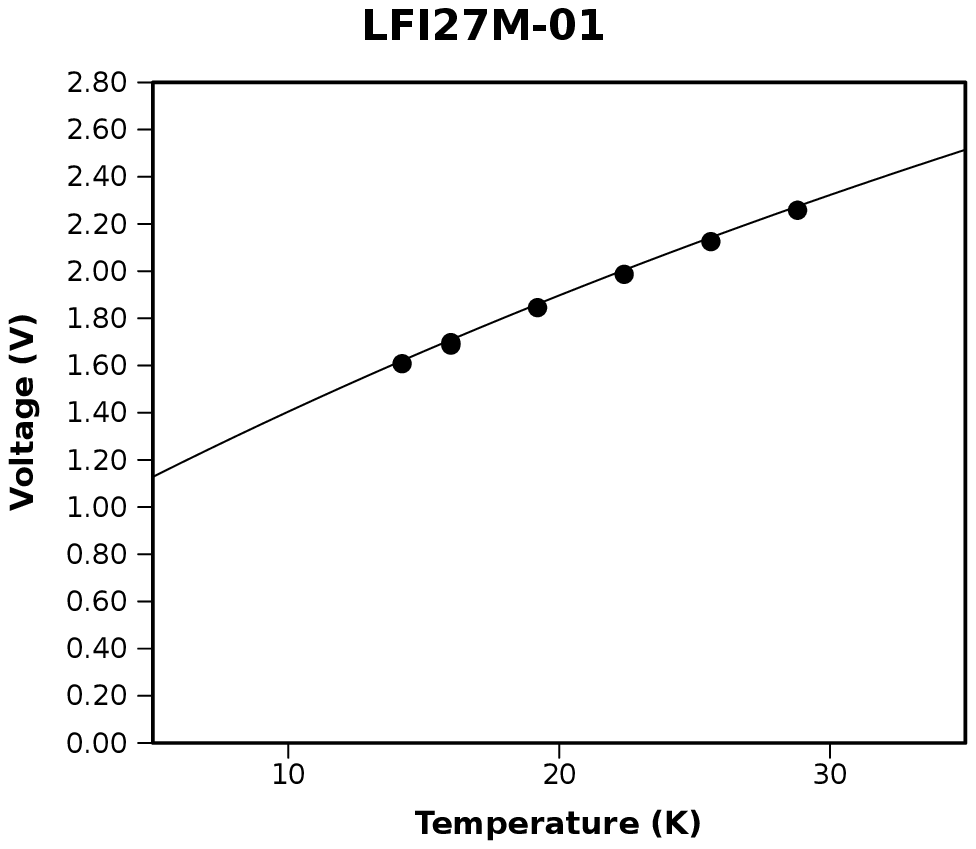}
        \includegraphics[width=3.5cm]{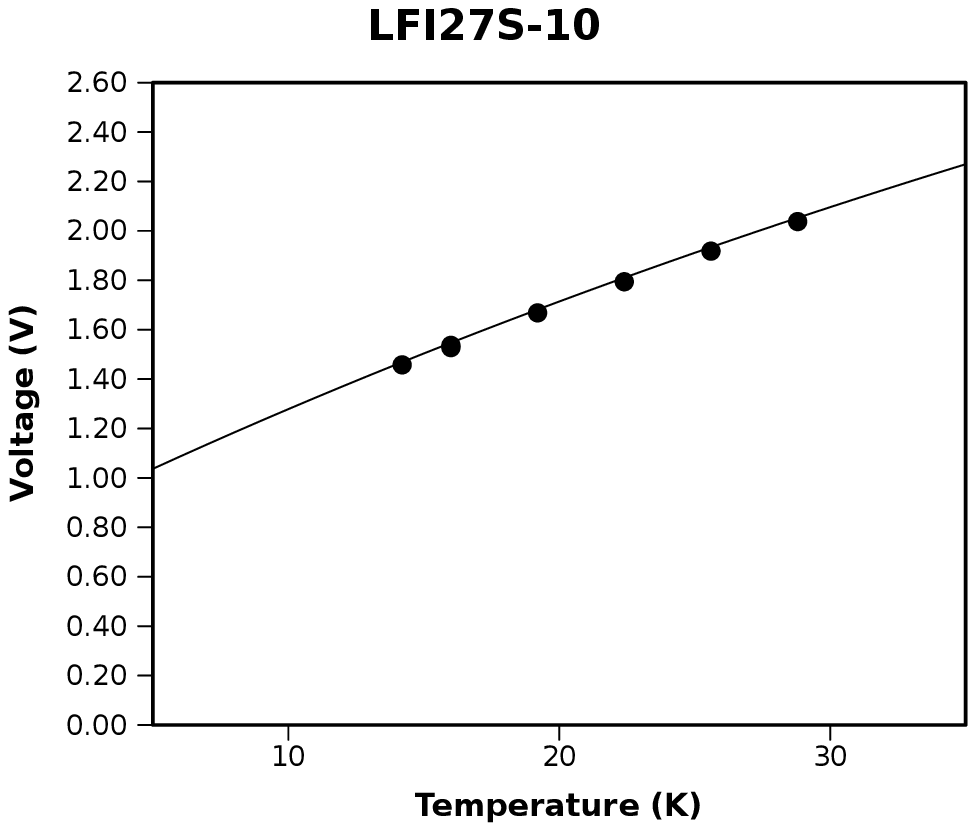}
        \includegraphics[width=3.5cm]{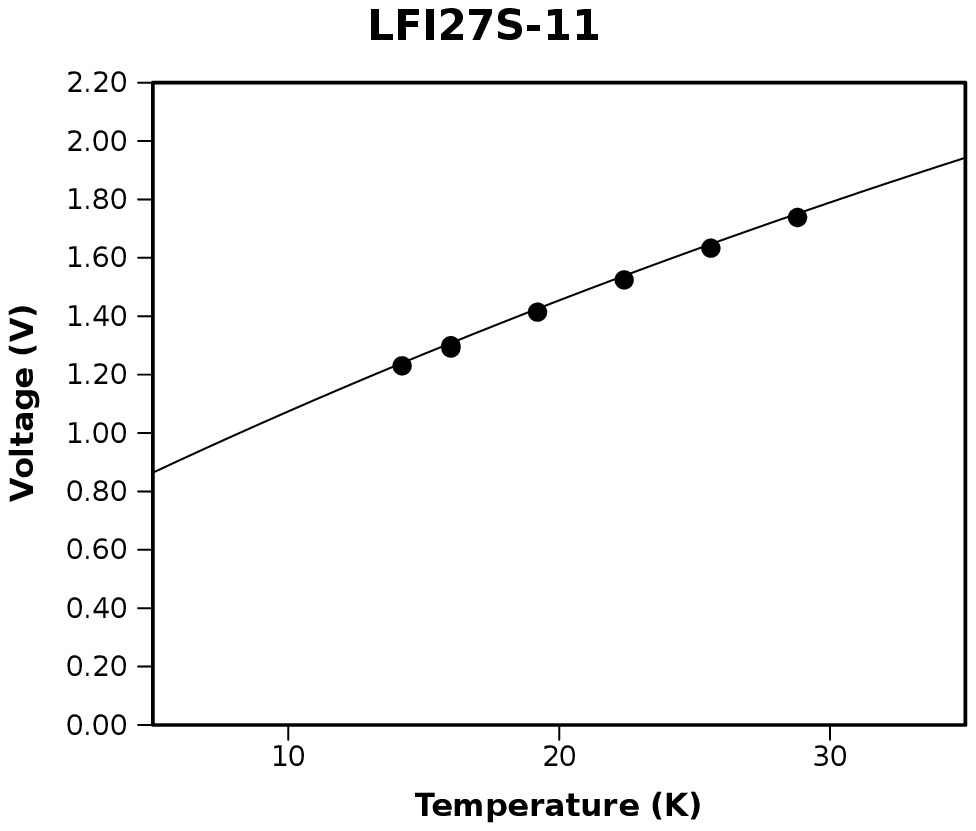}\\

        \includegraphics[width=3.5cm]{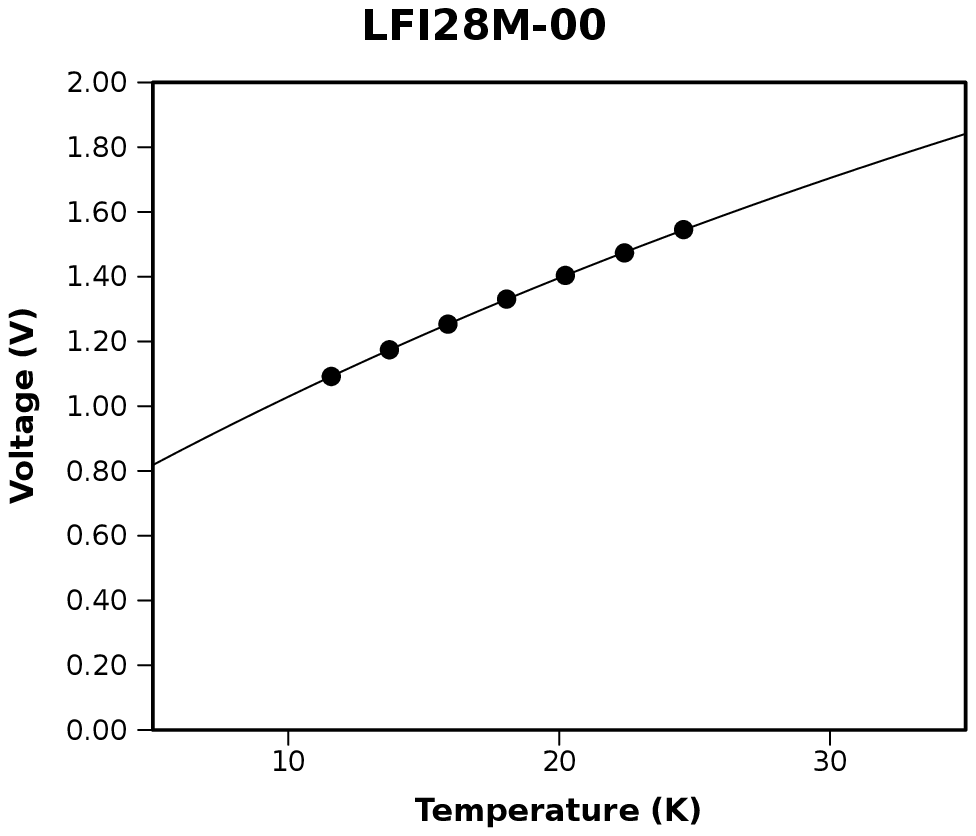}
        \includegraphics[width=3.5cm]{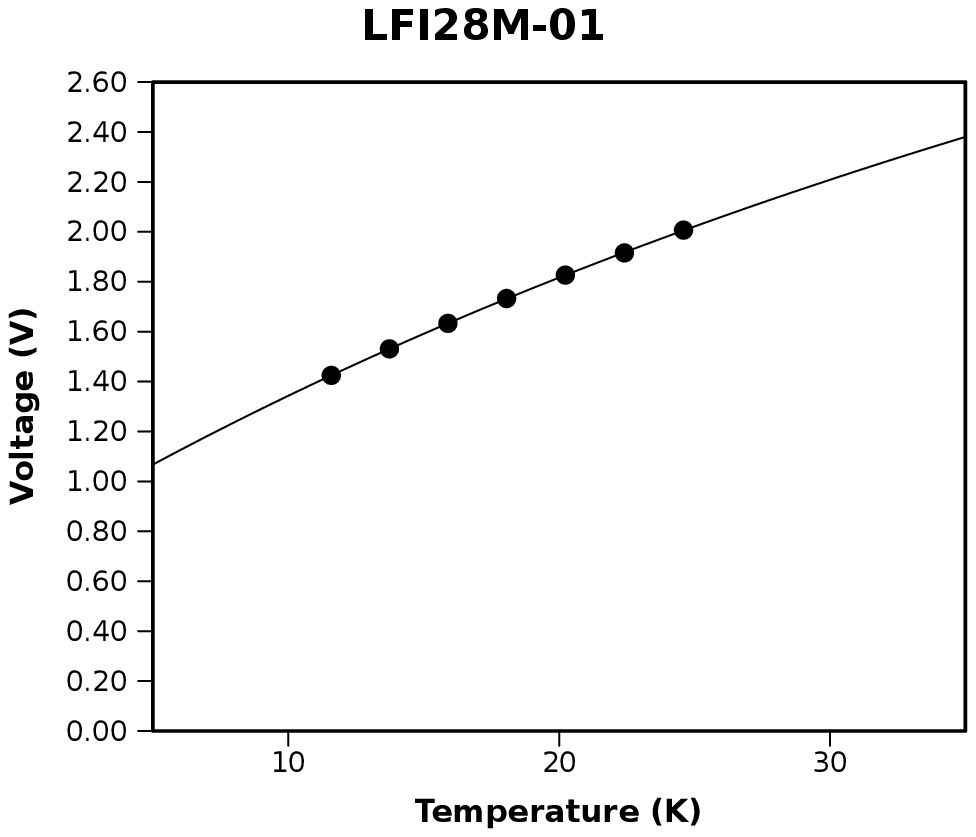}
        \includegraphics[width=3.5cm]{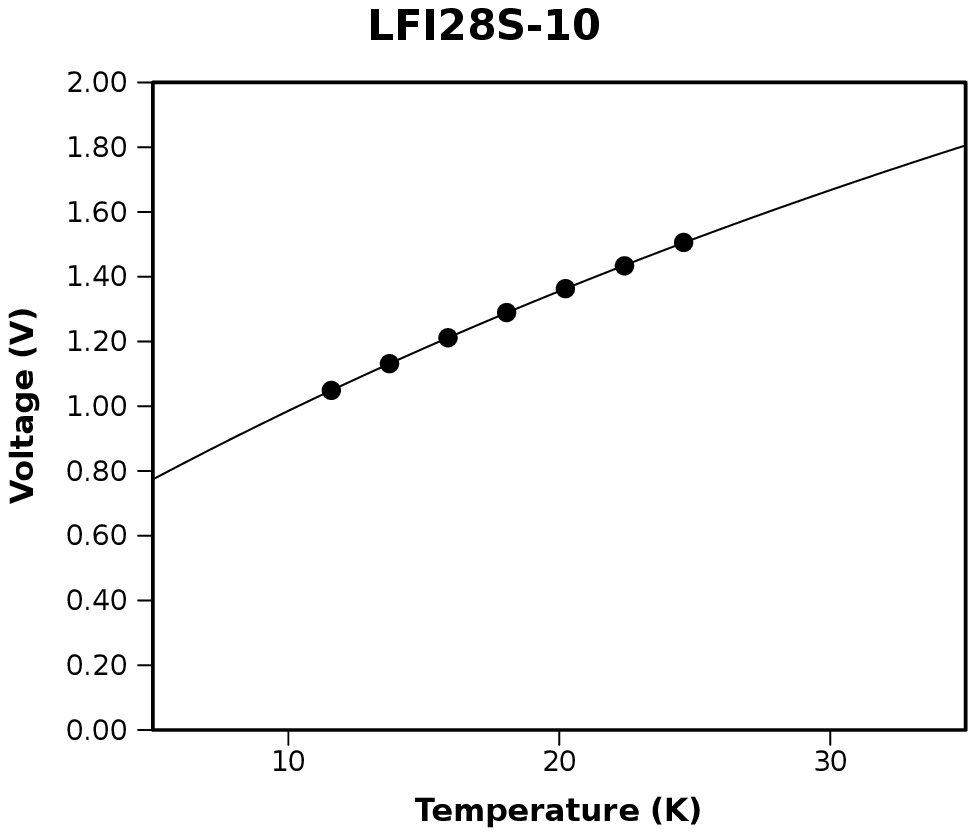}
        \includegraphics[width=3.5cm]{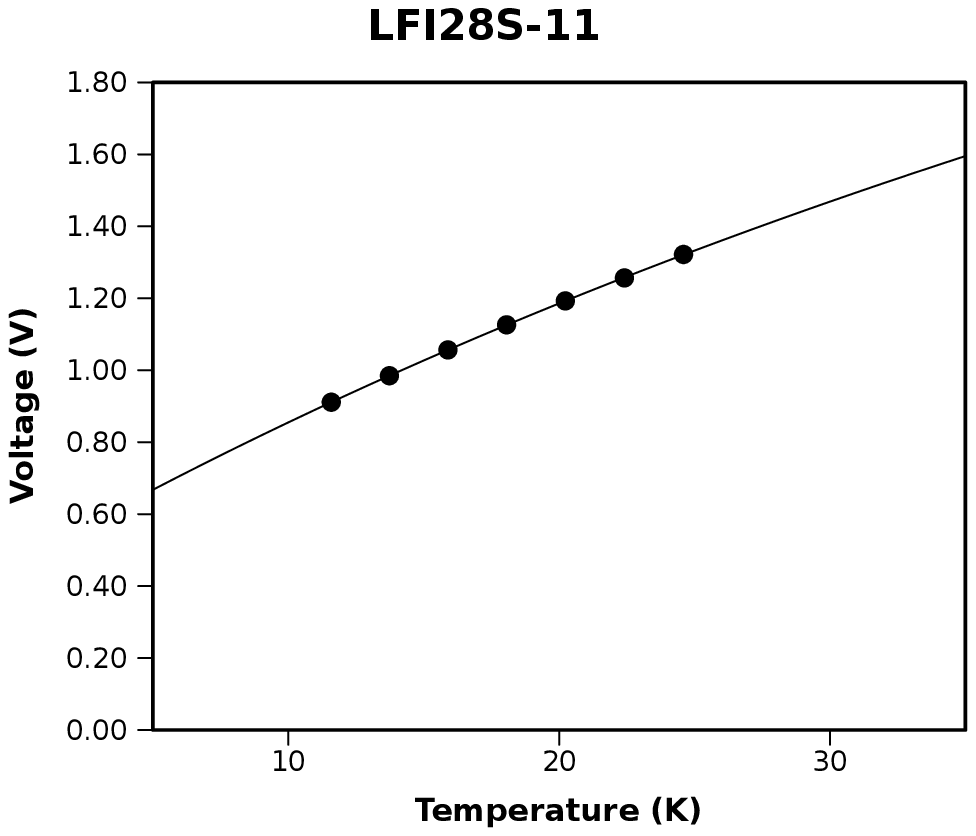}

    \end{center}
    \caption{Non linear fits for all the 30 and 44~GHz detectors with the parameters in Table~\protect \ref{tab:best_fit_parameters}}
\end{figure}




%
\bibliographystyle{JHEP}
\bibliography{R04_mennella}
\end{document}